\documentclass[oneside]{aastex}
\begin{document}

\title{Konus catalog of SGR activity to 2000}

\author{R.L. Aptekar', D.D. Frederiks, S.V. Golenetskii, V.N. Il'inskii,
E.P. Mazets, V.D. Pal'shin}
\affil{Ioffe Physico-Technical Institute, Russian Academy of Science}
\affil{St.Petersburg, 194021, Russia}
\author{P.S. Butterworth and T.L.Cline}
\affil{NASA Goddard Space Flight Center, Greenbelt, Maryland, 20771, USA}

\begin{abstract}
Observational data on the bursting activity of all five known Soft Gamma
Repeaters are presented. This information was obtained with Konus gamma-ray 
burst experiments on board Venera 11-14, Wind, and Kosmos-2326 spacecraft
in the period from 1978 to 2000. These data on appearance rates, 
time histories, and energy spectra of repeated soft bursts
obtained with similar instruments and collected together in a comparable
form should be useful for further studies of SGRs.
\end{abstract}

\section{INTRODUCTION}

Recurrent short gamma-ray bursts with soft energy spectra have been known
for twenty years. The first two sources of repeating bursts were discovered
and localized in March 1979 with the Konus experiment aboard the Venera 11 and 
Venera 12 missions (Mazets \& Golenetskii 1981).
The extraordinary intense gamma-ray outburst on 1979 March 5 (Mazets et al.
1979a) was followed by a series of 16 weaker short bursts from source FXP0526-66,
during the next few years (Golenetskii et al. 1984). At the same time, in
March 1979, three short bursts were detected from another source, B1900+14
(Mazets, Golenetskii, \& Guryan 1979). It was suggested that repeated soft
bursts represent a distinct class of events different in their
origin from the majority of gamma-ray bursts (Mazets \& Golenetskii 1981;
Mazets et al. 1982). In 1983, the Prognoz 9 and ICE spacecraft observed a numerous
series of soft recurrent bursts from the third source, 1806-20
(Atteia et al. 1987; Laros et al. 1987). The sources of recurrent soft bursts
have became known as soft gamma repeaters, SGRs.

Curiously, a retrospective analysis of Venera 11 and Prognoz-9 data showed
that the short burst of January 7, 1979 (Mazets et al. 1981) belonged to SGR~1806-20 
(Laros et al. 1986). Thus, the first three SGRs were detected within only three months.
However, as has become clear, SGRs are actually a very rare class of astrophysical
objects (Norris et al. 1991; Kouveliotou et al. 1994; Hurley et al. 1994). 
Indeed, the fourth source SGR~1627-41 was discovered and localized only in 1998 
(Hurley et al. 1999a; Woods et al. 1999; Smith et al. 1999; Mazets et al. 1999a). 
In 1997 two bursts were observed coming from the fifth SGR to be identified, 
SGR~1801-23 (Hurley et al. 1997; Cline et al. 2000).

The five known SGRs have displayed different levels of activity.
SGR~0526-66 has been silent since 1983. SGR~1900+14 emitted three bursts in
1979 (Mazets, Golenetskii, \& Guryan 1979) and three bursts in 1992 
(Kouveliotou et al. 1993). After a long period of silence,
the burst activity of SGR~1900+14 resumed in May 1998 at a very high and
irregular level and lasted until January 1999 
(Hurley et al. 1998; Hurley et al. 1999b; Mazets et al. 1999b).
On 1998 August 27 SGR~1900+14 emitted a giant outburst with a complex time
history which exhibited a striking similarity to the 1979 March 5 event
(Cline, Mazets, \& Golenetskii 1998; Mazets et al. 1999c; Hurley et al. 1999c).
SGR~1806-20 was a prolific source in 1979-1998 both in terms of active periods 
and in the numbers of emitted bursts (Laros et al. 1987; Atteia et
al. 1987; Kouveliotou et al. 1994; Frederiks et al. 1997; Marsden et al.
1997; G\"o\v{g}\"us et al. 2000). SGR~1627-41 was active in June-July 1998 
(Hurley et al. 1999a; Woods et al. 1999a; Mazets et al. 1999a). 
Only two bursts from SGR~1801-23 were detected, on June 29, 1997 (Cline et al. 2000).

X-ray studies of SGR~0526-66 (Rothschild et al. 1994),
SGR~1806-20 (Cooke 1993), SGR~1900+14 (Hurley et al. 1996),
and SGR~1627-41 (Woods et al. 1999a) have revealed quiescent soft X-ray
counterparts to all these objects. Measurements of a rapid spin-down
of SGR~1806-20 and SGR~1900+14 have been interpreted (Kouveliotou et al
1998; 1999) as establishing SGRs as ``magnetars",
i.e. as young isolated neutron stars with a superstrong magnetic field up to
10$^{14}$-10$^{15}$~G (Duncan \& Thompson 1992; Thompson \& Duncan 1995).
At the same time, arguments for significantly lower magnetic fields in SGRs
have also been made (Marsden, Rothschild, \& Lingenfelter 1999;
Harding, Contopoulos, \& Kazanas 1999).

This catalog contains data on soft bursts from five soft gamma repeaters
observed with the Konus experiments aboard the Venera 11, 12, 13, 14 missions 
in 1978-1983, with the Konus-Wind experiment on the Wind spacecraft in 1994-1999, 
and with the Konus-A experiment aboard Kosmos-2326 in 1995-1997.
Our data on time histories and energy spectra of bursts presented here
should be useful for further SGR studies especially when compared with
results from similar instruments.
The catalog including initial processed data is also available electronically at: 
http//www.ioffe.rssi.ru/LEA/SGR/Catalog/

\section{INSTRUMENTS AND OBSERVATIONS}

\subsection{The Konus experiment on the Venera missions}

The Konus instrument consisted of six identical gamma-ray detectors. Each
sensor had a NaI(Tl) scintillator 80 mm in diameter and 30 mm in height viewed by a
PM-tube through a thick lead glass window. An additional passive shield of a lateral
surface of the crystal resulted in a cosine-like angular sensitivity of the detector.
The axes of the six detectors were aligned along orthogonal axes of the spacecraft.
This arrangement provided an instrument capability for burst localization with an 
accuracy of between one and a few degrees depending on the intensity of the burst 
(Golenetskii, Il'inskii, \& Mazets 1974).
The instrument operated in a triggered mode.
The trigger threshold was set to be about $6\sigma$ above the current background level
in the energy window 50-150 keV. When triggered, the instrument recorded a burst time
history with resolutions of 1/64, 1/4, and 1~s as well as 8 energy spectra in the energy
region 30~keV--2~MeV measured with an accumulation time of 4~s (Mazets at al. 1979b).

Observations of gamma-ray bursts (GRBs) from Veneras 11 and 12
were made from September 1978 until February 1980. The experiment was
continued on Veneras 13 and 14 from November 1981 to April 1983 with improved spectral
and time resolution (Mazets et al. 1983).

\subsection{The Konus-Wind experiment}

On the U.S. GGS Wind spacecraft, two scintillation gamma-ray detectors are
monitoring northern and southern ecliptic hemispheres. Each detector
contains a NaI(Tl) crystal 130 mm in diameter and 75 mm in height
in a housing with an entrance window made of beryllium. A burst time history is recorded
in three energy windows G1(10-50 keV), G2(50-200 keV), G3(200-750 keV) with a
variable time resolution from 2~ms up to 256~ms. These records include prehistory sections
of event time profiles recorded with 2~ms time resolution. Up to 64 energy spectra are
measured in the energy range 10~keV--10~MeV. The accumulation time for each spectrum is
automatically adapted to the current burst intensity within the range from 64~ms to 8~s.
Time history records are of fixed duration 229.632 s. Spectral measurement
may take significantly shorter or longer depending on the total
accumulation time for the set of 64 energy spectra (Aptekar et al. 1995).
Observations of GRBs from Wind have been made since November 1994.

\subsection{The Konus-A experiment}

The near-Earth orbiting Kosmos-2326 spacecraft was equipped with the Konus-A
experiment for the study of GRBs. This experiment consisted of two gamma-ray burst
detectors with associated electronic units. The first was a spectroscopic
gamma-ray detector identical to the burst detectors flown on the Wind spacecraft.
The second was an ensemble of four directionally sensitive gamma-ray detectors which
were arranged to create an instrumental capability to localize a burst arriving from
the zenith-centered hemisphere. Studying GRBs by means of two identical instruments 
from two spacecraft increased significantly the reliability of identifying faint
details and features in burst time histories and energy spectra.
Observations from Kosmos-2326 were carried out from December 1995 to October 1997.

\section{CATALOG}

\subsection{Catalog structure}

The catalog presents data from three experiments in chronological order but
separately for each SGR. The information for each SGR is displayed in the same sequence.

First, a Table is given which contains the main characteristics of the
events. The first  three columns specify burst order numbers, burst names 
according to date of appearance, and trigger times T$_0$. Burst duration $\Delta$T
is in the fourth column. Burst rise and fall times are both determined using the
criterion that a burst was present only while the observed counts
exceeded the background level by over three $\sigma$ in three successive time bins
in the time history. The next two columns present values of peak fluxes
and fluences. The spectral parameter kT is given in the seventh column.
To ensure uniformity of data presentation, the kT values were obtained by
fitting energy photon spectra to optically-thin thermal bremsstralung using:
\begin{equation}
F(E_\gamma) \propto E_\gamma^{-1} \exp\left(-\frac{E_\gamma}{kT}\right).
\end{equation}
The last column contains remarks related mainly
to measurements on the Wind and Kosmos spacecraft. Events
detected in the trigger mode are marked with an index `T'. They provide the
most complete temporal and spectral data. 
In some cases more than one event can be detected during
a trigger mode record which lasts about four minutes.
These events are marked with index `S'. The values T$_0$ shown 
in the third column in these cases denote the time of appearance of a serial event.
The mark `B' indicates weaker events observed in the background mode which
provides a coarse time resolution and no direct spectral data. In this case,
the kT value was evaluated from the accumulated ``hardness ratio" G2/G1.
For the weakest events, only an upper limit for kT could be obtained. 
Only coarse G2 count rate values are available in the
spacecraft houskeeping data for events that occured during the ``dead time"
of the instrument, when information about previous triggered events is read out.
Such events are denoted by the mark `H'.

Second, a set of figures is presented showing time histories and energy
spectra for the observed bursts. Time histories are usually given for the 
two energy windows G1 and G2 together with the count rate ratio G2/G1.
Deconvolved background-subtracted photon spectra were
accumulated in the proper time intervals. On the time history graphs, these
intervals are shown enclosed between two vertical dotted lines. 
For some weak events, an important fraction of the burst falls into 
the prehistory period. A spectrum accumulated after T$_0$ is
often too uncertain to be shown on the graph. In this case, the parameter kT
must be evaluated from
the hardness ratio G2/G1 integrated for the whole burst.
In some cases e.g. for serial events, a spectrum accumulating time is much
longer that burst duration. Photon spectra averaged formally over an accumulation 
time interval and shown in Figures exhibit strongly reduced intensity.
If the burst duration is known the flux can be easy
corrected for the given $\Delta$T.

Finally, there are several special events the study of which is expected to
be of great importance for a deeper understanding of the SGR origin. More extensive
information concerning these outbursts and event series is presented together with
comments and explanations.

The data presented have been corrected for interferences, dead
time, and all other known distortions. After background subtraction,
instrumental energy-loss spectra were deconvolved to incident photon spectra.

\subsection{SGR~0526-66}

This is the first known source of soft repeated bursts. It was discovered
and localized with the Konus experiment on board the Venera~11 and Venera~12 missions in
1979 when on March 5 a giant outburst accompanied by a long pulsating (8~s) tail was observed
to be followed a few days later by several weak recurrent bursts (Mazets et al. 1979a). A much
more precise localization performed by Cline et al. (1982) resulted in a very small error
box which projected onto the supernova remnant N~49 located in the Large Magellanic
Cloud (LMC) at a distance of 55~kpc.
This source continued moderate activity until 1983 (Golenetskii et al.
1984). A weak persistent soft X-ray flux from SGR~0526-66
in its quiescent state has been detected with ROSAT (Rothschild et al. 1994). 
The data for 17 events are presented in Table~2.1 and displayed in Figures 2.1--2.18.

\subsection{SGR~1900+14}

The second source of soft recurrent bursts was also discovered and localized
in March 1979 (Mazets, Golenetskii \& Guryan 1979). 
Several bursts from this SGR were observed during its reactivation
period in 1992 by BATSE (Kouveliotou et al. 1993). SGR~1900+14 is believed
to be associated with the supernova remnant G42.8+0.6 (Kouveliotou et al. 1994)
situated at distance of 10.4~kpc (Case \& Bhattacharya 1998) but see also
Vrba et al. (2000).
ASCA observations in April 1998 revealed a 5.16~s periodicity in persistent X-ray
emission from this source (Hurley et al. 1999d). A further episode of bursting activity began
in May 1998 and continued until January 1999. BATSE (G\"o\v{g}\"us et al. 1999), BeppoSAX
(Feroci et al. 1999), and Konus-Wind (Mazets et al. 1999b) investigated numerous bursts.

On August 27, 1998, a giant outburst in SGR~1900+14 was observed
(Cline, Mazets, \& Golenetskii 1998; Hurley et al. 1999c; Feroci et al. 1999).
This event strikingly resembled the March 5 outburst from
SGR~0526-66. It also consisted of a huge brief initial pulse followed by a
slowly decaying and coherently pulsating (5.16~s) tail (Mazets et al. 1999c). 
The close similarity between these two giant outbursts indicates their common nature.

The appearance rate of repeated bursts was very irregular and clusters of
bursts were observed. During the most crowded cluster, events followed one after
another in time intervals as short as a few burst lengths.

The data related to SGR~1900+14 are presented in Table~3.1 and shown in
Fig.~3.1--3.66.

\subsection{SGR~1806-20}

The third repeater, SGR~1806-20 was discovered and localized during a period
of high activity in 1979-1983 in observations from the ICE and Prognoz-9 spacecraft
(Laros et al. 1986; Laros et al. 1987; Atteia et al. 1987).

In subsequent years, the source continued a moderate level of activity.
Beginning in 1996, the source became very active again with many clusters of events.
A persistent soft X-ray source was observed by Murakami et al. (1994). 
Kouveliotou et al. (1998) discovered 7.47~s pulsations in
the X-ray flux. Kulkarni and Frail (1993) located SGR~1806-20 inside the
supernova remnant G10.0-0.3 at a distance of $\sim14$~kpc. Statistical properties
of the repeating bursts were studied by G\"o\v{g}\"us et al. (2000).

Results of observations of SGR~1806-20 with Konus experiments are presented
in Table~4.1 and Fig.~4.1--4.20.

\subsection{SGR~1627-41}

This source was discovered in summer 1998 with observations from CGRO (Woods
et al. 1999), Ulysses (Hurley et al. 1999a), Wind (Mazets et al. 1999a), and RXTE
(Smith et al. 1999). The source was precisely localized by the IPN. 
Its position coincides with the supernova remnant G337.0-0.1 at a distance 
of $\sim11$~kpc (Hurley et al. 1999a). Some evidence was obtained in support of a
possible periodicity of 6.7~s (Dieters et al. 1998; Woods et al. 1999), but
the observations from ASCA didn't confirm this result (Hurley et al. 2000).

An enormously intense outburst from this source was observed on 1998 June
18. It was a short single pulse without any evidences for the existence of a 
pulsating decay (Mazets et al. 1999a).
Calculation of the outburst intensity has required major dead time
corrections.
It is of great interest that on June 18 at least one more such event occurred.
However, it was detected only in housekeeping data. This circumstance prevented  
us from making reliable dead time corrections. Hence, only a lower limit of
intensity for this burst could be obtained.

Our data from the Konus-Wind experiment are presented in Table~5.1 and
Fig.~5.1--5.17.

\subsection{SGR~1801-23}

Only two short bursts from this source were detected on June 29, 1997 by
CGRO, Ulysses, Wind and Kosmos-2326 (Cline et al. 2000). 
However, it should be remembered that the first active period
of SGR~1900+14 observed in March 1979 consisted of only three
events. An association of SGR~1801-23 with the supernova remnant G6.4-0.1
is possible.

The data obtained are presented in Table~6.1 and Fig.~6.1.

\section{CONCLUSION}

This catalog confirms the existence of many similarities between
the five known SGRs.
The most obvious are a short duration of repeated bursts
and their soft energy spectra.
The two giant outbursts are strikingly similar.
Moreover, all SGRs appear to be associated with young
supernova remnants. Weak persistent soft X-ray fluxes
were detected from four SGRs in a quiescent state.
A regular periodicity of 5-8~s was well established
for three SGRs in the X-ray and/or gamma-ray energy ranges.
A fast spin-down  of $10^{-10}$~ss$^{-1}$ was determined
for SGR~1806-20 and SGR~1900+14.
The energy release in SGRs if they are at distances of $\sim10$~kpc averages
for repeated bursts $\sim10^{40}-10^{41}$~erg and for giant
outbursts $\sim10^{44}$~erg.
Correspondingly, the luminosity of SGRs exceeds the Eddington limit by
factors of thousands and millions respectively.
All these common features have provided an observational basis
for the magnetar model of SGRs.
At the same time, some distinctions between the properties
of repeated bursts can also be seen in the data
presented in this catalog.
For example, note the different patterns of spectral
variability for SGR~1627-41, SGR~1900+14, and SGR~1806-20.

Studying such similarities, distinctions,
and individual features should lead to a deeper understanding of the
fleeting but extremely  powerful processes operating in SGRs.

This work was supported by Russian Aviation and Space Agency Contract
and RFBR grant N~99-02-17031.

\setcounter{subsection}{2}
\setcounter{table}{0}
\begin{deluxetable}{llccccccc}
\tabletypesize{\scriptsize}
\tablecaption{SGR 0526-66\tablenotemark{a}}
\tablehead{
\colhead{N} & \colhead{Burst} & \colhead{T$_0$}& \colhead{$\Delta$T} & \colhead{P$_{max}$} &%
\colhead{S} & \colhead{kT} & \colhead{Comments} & \colhead{Figures}\\
& \colhead{name} & \colhead{h:m:s UT}& \colhead{(s)}&\colhead{(erg cm$^{-2}$ s$^{-1}$)} &%
\colhead{(erg cm$^{-2}$)} & \colhead{(keV)}&  &\colhead{PNG file}
}
\startdata
1 & 790305 & 15:51:39.145 & & & & & T & 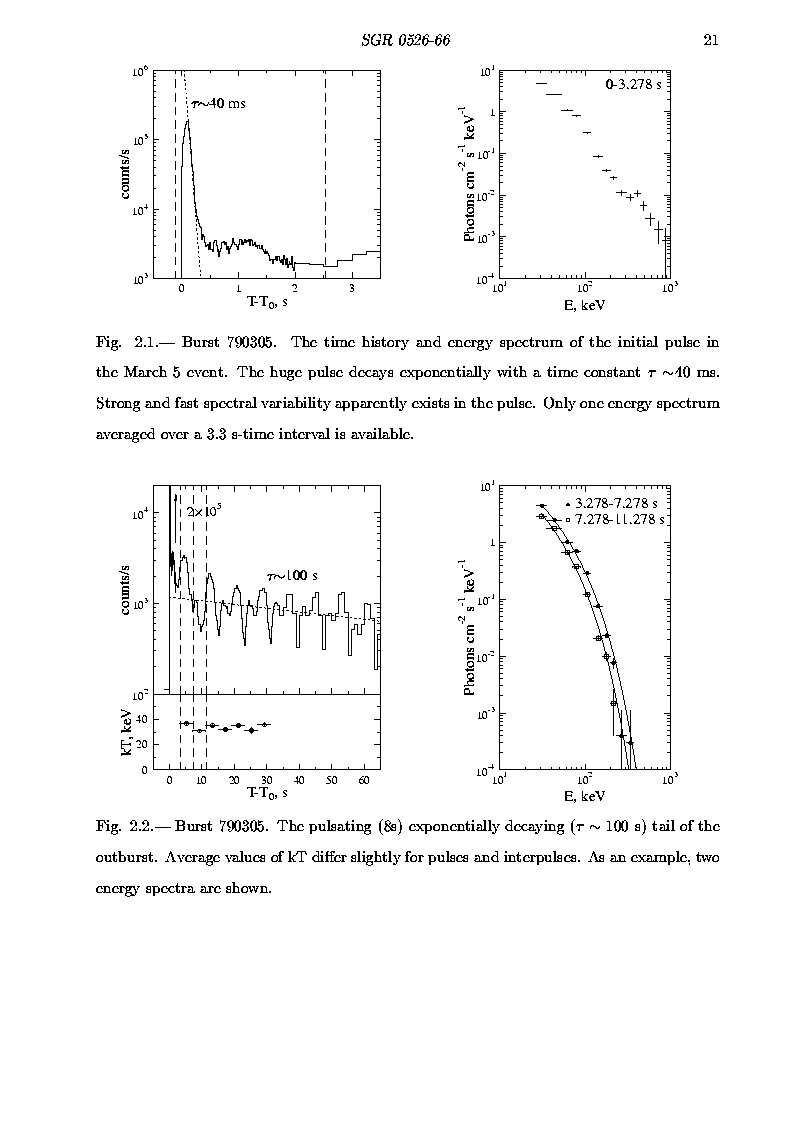 \\
 & Initial pulse &  & 0.25 & 1.0$\times 10^{-3}$ & 4.5$\times 10^{-4}$ & 500$\to$30 & &  \\ 
 & Pulsating tail & & $>64$ &  & 1.0$\times 10^{-3}$ & $\sim30$ & & \\ 
2 & 790306 & 06:17:25.201 & 1.5 & 4.4$\times 10^{-6}$ & 6.5$\times 10^{-6}$& 31$\pm 5$ & T & 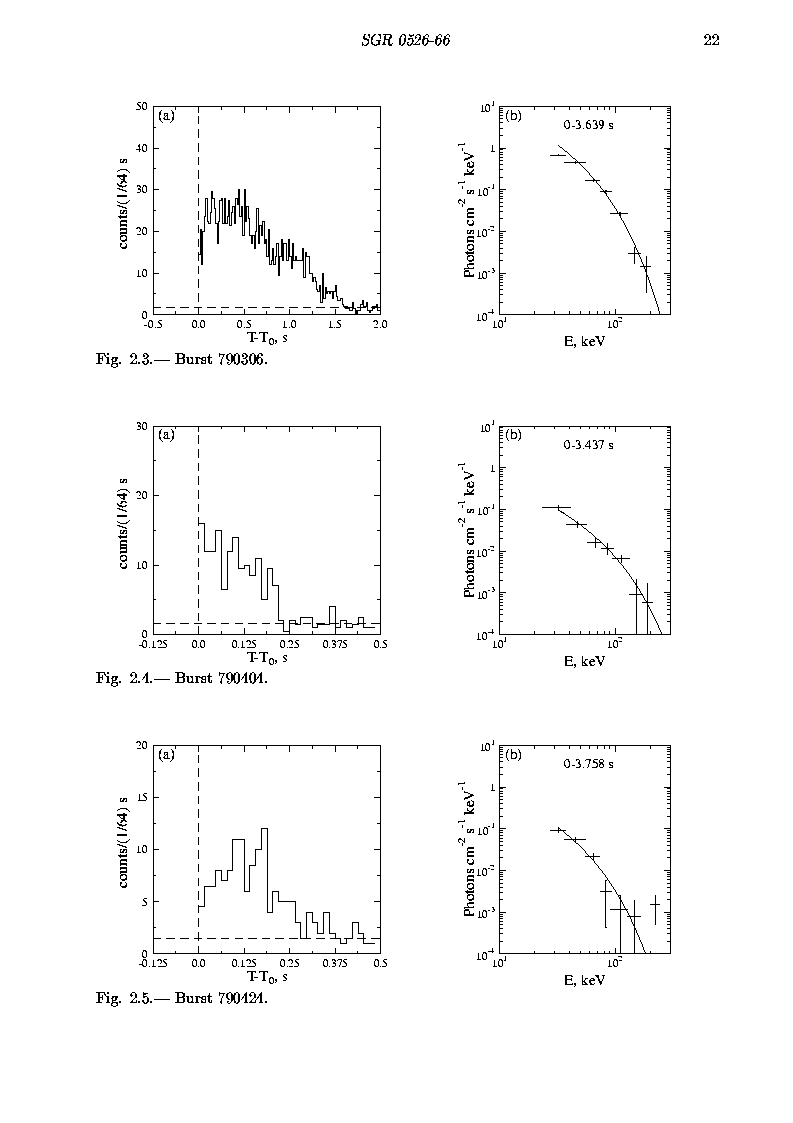 \\ 
3 & 790404 & 00:43:31.317 & 0.2 & 2.6$\times 10^{-6}$ & 7.0$\times 10^{-7}$& 30$\pm 10$ & T & SGRfig02.png \\ 
4 & 790424 & 18:27:18.241 & 0.2 & 2.1$\times 10^{-6}$ & 6.0$\times 10^{-7}$& 30$\pm 15$& T& SGRfig02.png \\ 
5 & 811201 & 09:18:02.424 & 3.5 & 2.1$\times 10^{-6}$ & 6.6$\times 10^{-6}$& 30$\pm 2$ & T& 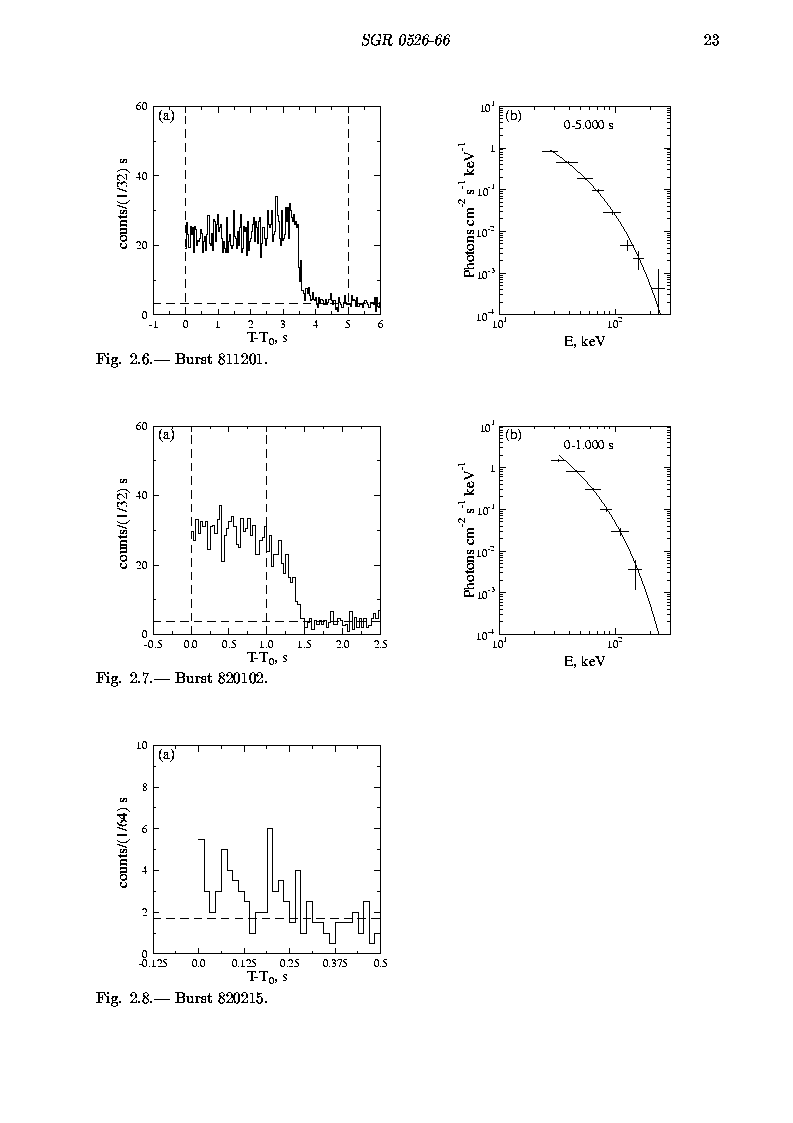 \\ 
6 & 820102 & 17:11:11.109 & 1.5 & 2.6$\times 10^{-6}$ & 3.6$\times 10^{-6}$& 26$\pm 3$ & T & SGRfig03.png \\ 
7 & 820215 & 10:36:31.071 & 0.1 & 1.2$\times 10^{-6}$ & 1.5$\times 10^{-7}$& 40$\pm 15$& T& SGRfig03.png \\ 
8 & 820227 & 00:16:41.178 & 9.0 & 7.2$\times 10^{-6}$ & 2.2$\times 10^{-5}$& 28$\pm 3$ & T & 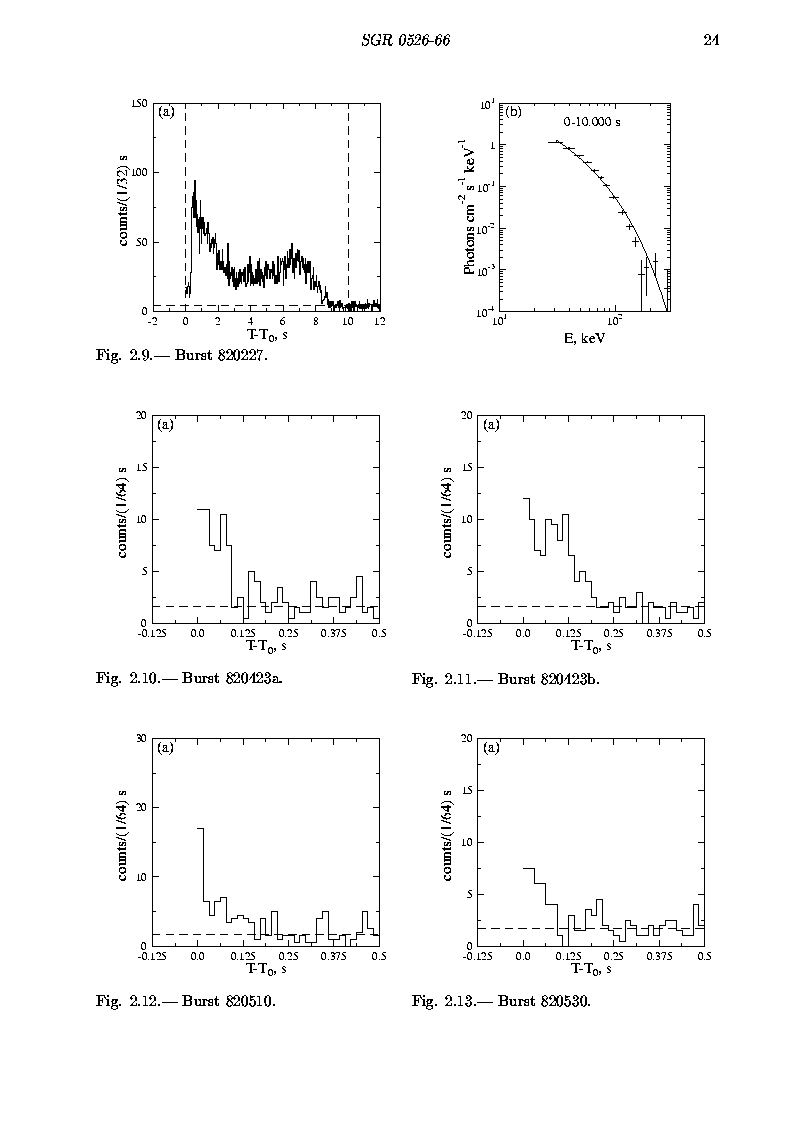 \\ 
9 & 820423a & 03:43:16.907 & 0.1 & 2.3$\times 10^{-6}$ & 2.0$\times 10^{-7}$& 36$\pm 10$ & T & SGRfig04.png \\ 
10 & 820423b & 16:29:21.315 & 0.15 & 2.6$\times 10^{-6}$ & 3.0$\times 10^{-7}$& 45$\pm 15$ & T & SGRfig04.png \\ 
11 & 820510 & 13:27:55.921 & 0.1 & 1.8$\times 10^{-6}$ & 1.5$\times 10^{-7}$& \nodata & T  & SGRfig04.png \\ 
12 & 820530 & 12:46:13.968 & 0.1 & 1.8$\times 10^{-6}$ & 1.5$\times 10^{-7}$& 35$\pm 15$ & T  & SGRfig04.png \\ 
13 & 820708 & 10:18:18.979 & 0.1 & 2.1$\times 10^{-6}$ & 1.5$\times 10^{-7}$& 30$\pm 15$ & T & 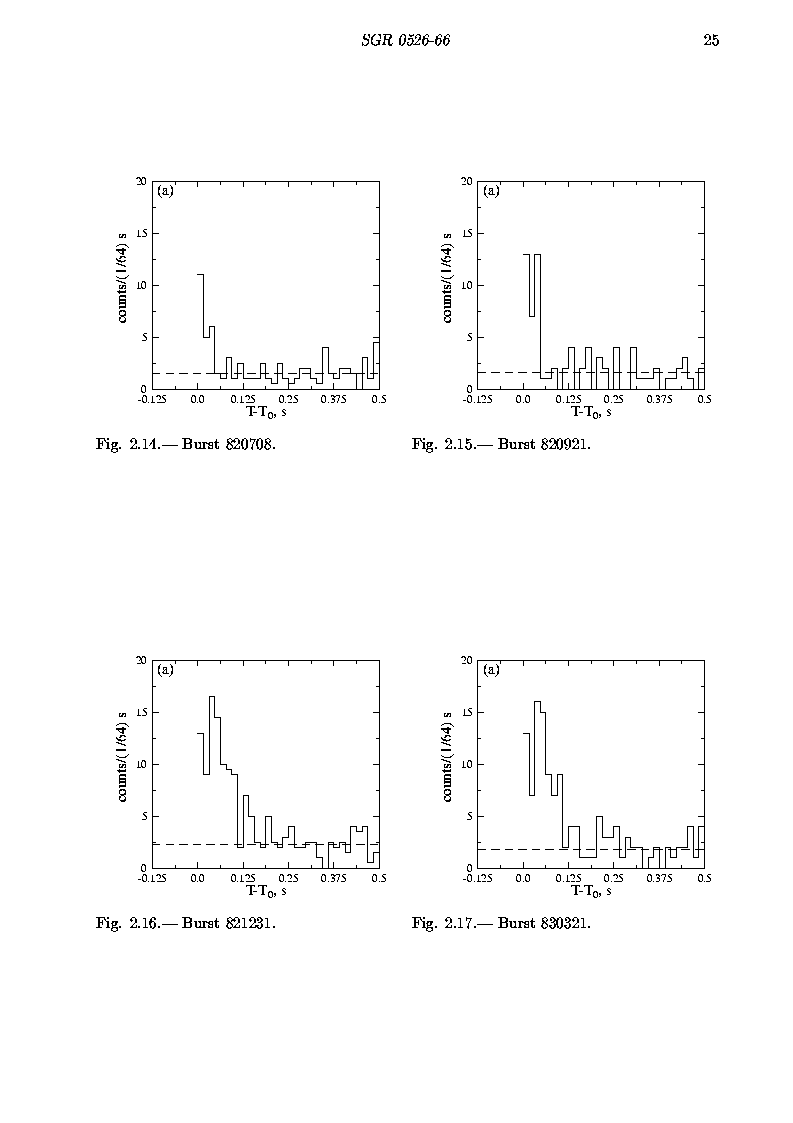 \\ 
14 & 820921 & 18:37:54.138 & 0.1 & 1.8$\times 10^{-6}$ & 1.5$\times 10^{-7}$& 27$\pm 15$ & T & SGRfig05.png \\ 
15 & 821231 & 04:17:21.363 & 0.18 & 2.5$\times 10^{-6}$ & 3.0$\times 10^{-7}$& 27$\pm 10$ & T & SGRfig05.png \\ 
16 & 830321 & 22:32:11.400 & 0.15 & 2.1$\times 10^{-6}$ & 2.0$\times 10^{-7}$& 24$\pm 10$ & T& SGRfig05.png \\ 
17 & 830405 & 07:58:07.667 & 1.1 & 2.6$\times 10^{-6}$ & 1.6$\times 10^{-6}$& 31$\pm 6$ & T  & 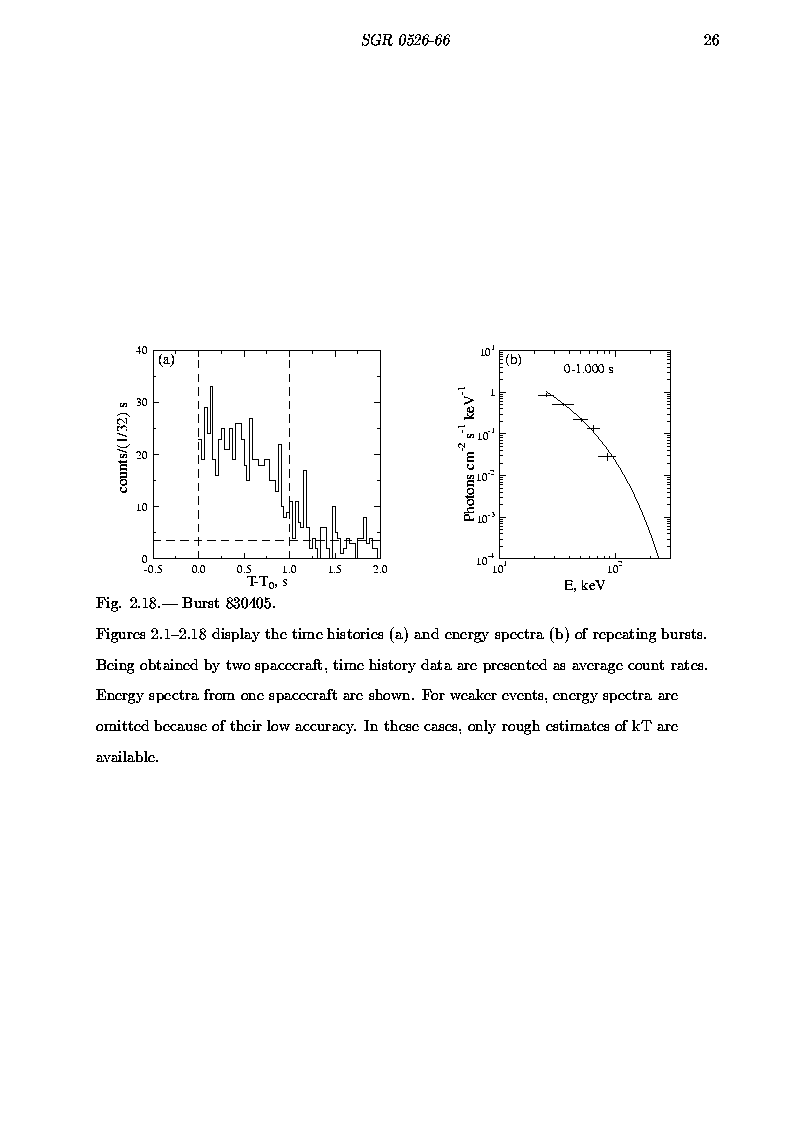 \\ 
\enddata
\tablenotetext{a}{The table begins with the March 5 event which is naturally divided into two
phases: the giant short initial pulse followed by a long pulsating (8 s) tail (see Fig. 1).
Data for these two phases are given separately. Following lines present
the data for repeated weaker bursts.}
\end{deluxetable}

\setcounter{subsection}{3}
\setcounter{table}{0}
\begin{deluxetable}{lllcccccc}
\tabletypesize{\scriptsize}
\tablecaption{SGR 1900+14\tablenotemark{a}}
\tablehead{
\colhead{N} & \colhead{Burst}& \colhead{T$_0$}& \colhead{$\Delta$T} & \colhead{P$_{max}$} &%
\colhead{S} & \colhead{kT} & \colhead{Comments} & \colhead{Figures}\\
& \colhead{name} &\colhead{h:m:s UT}& \colhead{(s)}&\colhead{(erg cm$^{-2}$ s$^{-1}$)} &%
\colhead{(erg cm$^{-2}$)} & \colhead{(keV)} & &\colhead{PNG file}
}
\startdata
1 & 790324  & 16:06:49.826 & 0.12 & 1.3$\times10^{-5}$ & 1.0$\times10^{-6}$ & 35$\pm2$& T & 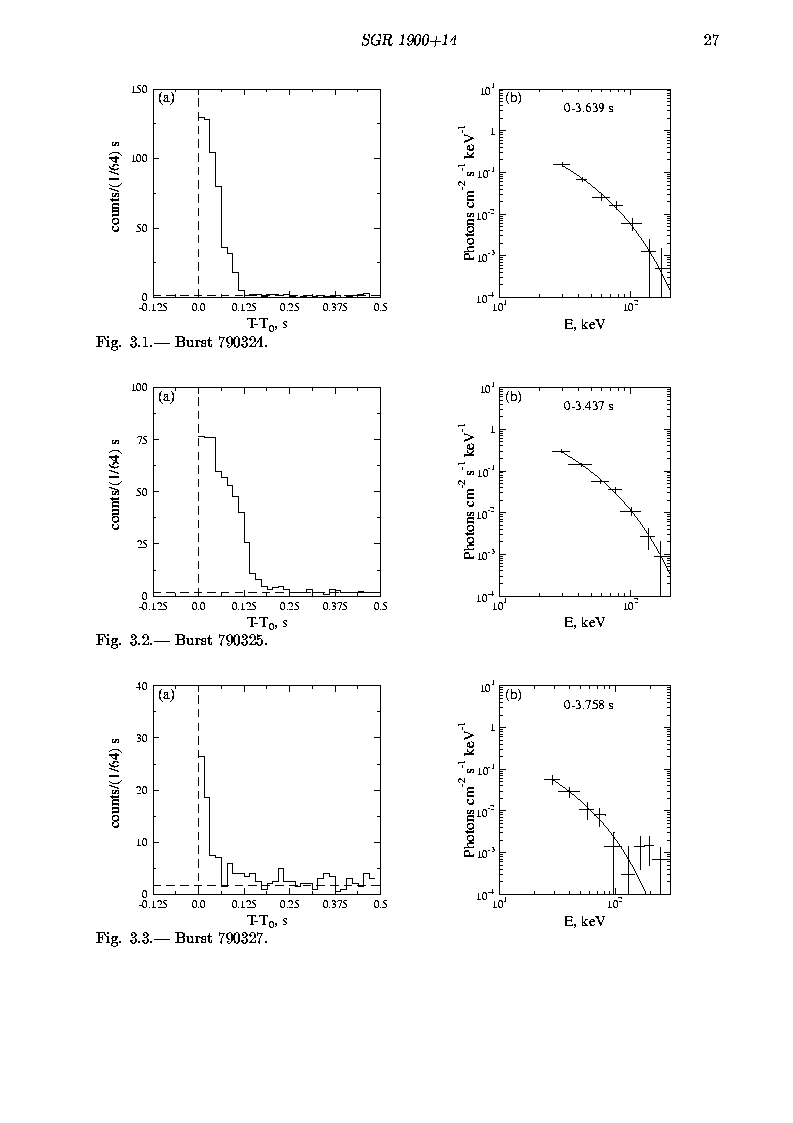 \\
2 & 790325  & 01:58:19.017 & 0.19 & 7.6$\times10^{-6}$& 1.5$\times10^{-6}$ & 36$\pm$2 & T & SGRfig07.png \\
3 & 790327  & 10:30:34.886 & 0.05 & 3.0$\times10^{-6}$ & 3.5$\times10^{-7}$ & 30$\pm15$ & T & SGRfig07.png \\
4 &980526  & 22:24:09.458	& 0.33 & 1.6$\times10^{-5}$ & 8.1$\times10^{-7}$ & 27$\pm$2 & T & 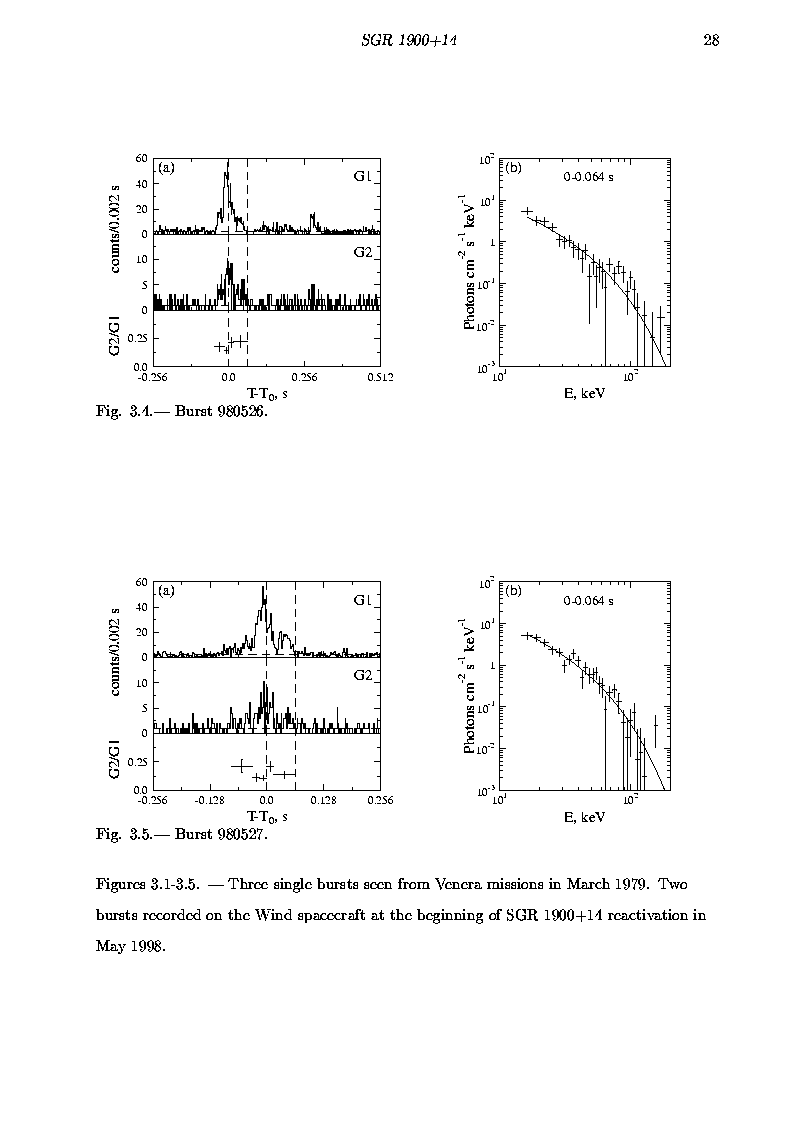 \\		
5 &980527  & 04:14:31.202   & 0.12 & 1.6$\times10^{-5}$ & 7.5$\times10^{-7}$ &	27$\pm$3 & T   & SGRfig08.png \\
6 &980530a  & 09:03:35.886	& 0.30 & 1.7$\times10^{-5}$ & 7.0$\times10^{-7}$ & 32$\pm$2 & TS & 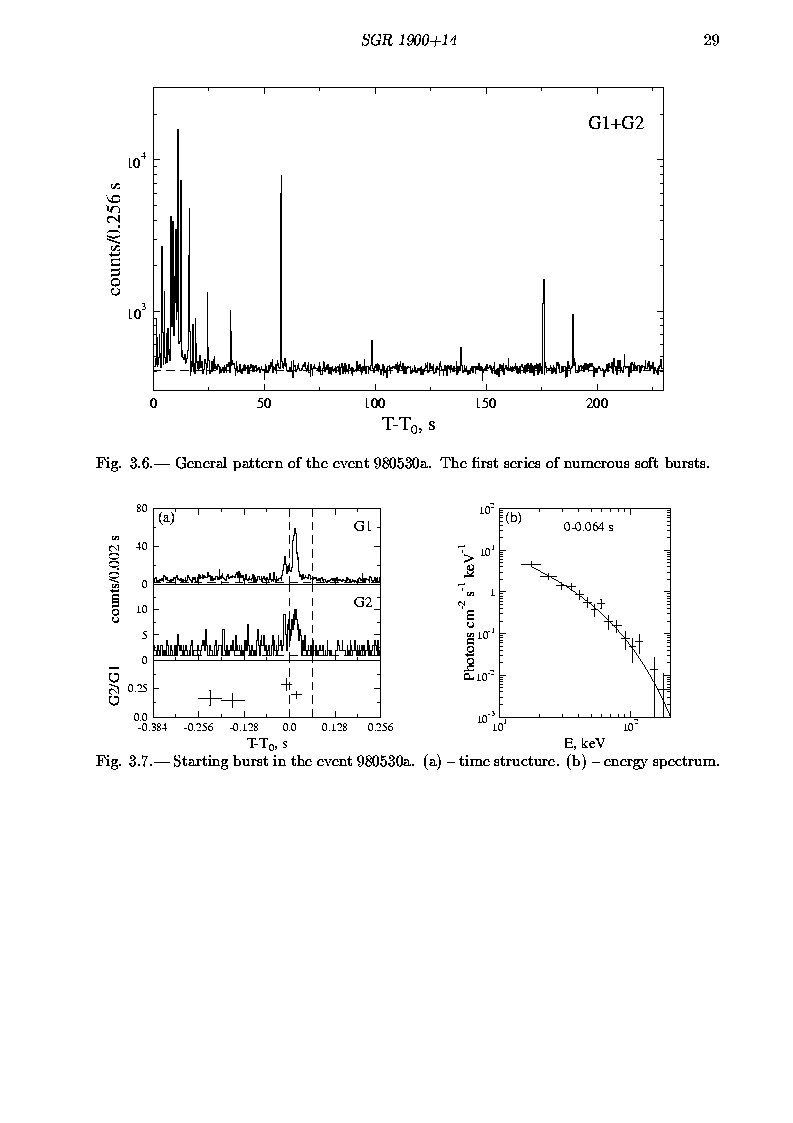 \\
 & & & & & &  & & --SGRfig13.png \\
7 & & 09:03:37.262 & 0.30 & 6.0$\times10^{-6}$ & 4.6$\times10^{-7}$ & 20$\pm$2 & S & \\
8 & & 09:03:38.366 & 0.42 & 1.9$\times10^{-6}$	& 2.8$\times10^{-7}$ & 22$\pm$4 & S  &  \\
9 & & 09:03:39.678 & 0.42  &	1.3$\times10^{-5}$	& 1.7$\times10^{-6}$ &	20$\pm$1 & S &  \\
10& & 09:03:40.670	& 0.27  &	1.4$\times10^{-5}$	& 6.4$\times10^{-7}$ &	28$\pm$2 & S  &  \\
11& & 09:03:41.886	& 0.18  &	1.4$\times10^{-6}$	& 1.8$\times10^{-7}$ &	30$\pm$5 & S &  \\
12& & 09:03:42.414	& 0.14  &	2.4$\times10^{-6}$	& 2.1$\times10^{-7}$ & 18$\pm$3 & S &  \\
13 & & 09:03:43.822	& 0.30  &	1.5$\times10^{-5}$	& 2.6$\times10^{-6}$ & 19$\pm$1 & S  &  \\
14 & & 09:03:44.350	& 0.22  &	5.0$\times10^{-6}$	& 3.7$\times10^{-7}$ &	18$\pm$3 &S &  \\
15 & & 09:03:44.654	& 0.37  &	1.6$\times10^{-5}$	& 3.0$\times10^{-6}$ &	20$\pm$1 & S &  \\
16 & & 09:03:45.470	& 0.22  &	1.1$\times10^{-5}$	& 9$\times10^{-7}$ &   29$\pm$1 & S &  \\
17 & & 09:03:45.950	& 0.19  &	2.0$\times10^{-5}$	& 2.0$\times10^{-6}$ &	26$\pm$1 & S &  \\
18 & & 09:03:46.782& 0.60  &	4.9$\times10^{-5}$	& 1.4$\times10^{-5}$ &	26$\pm$1 & S &  \\
19 & & 09:03:48.046	& 0.54  &	3.6$\times10^{-5}$	& 7.8$\times10^{-6}$ &	21$\pm$1 & S &  \\
20 & & 09:03:51.550	& 0.08  &	9.6$\times10^{-6}$	& 4$\times10^{-7}$ &	27$\pm$2 & S &  \\
21 & & 09:03:51.774	& 0.50  &	2.4$\times10^{-5}$	& 4.0$\times10^{-6}$ &	27$\pm$1 & S &  \\
22 & & 09:03:52.382 & 0.08  &	4.1$\times10^{-6}$	& 1.5$\times10^{-7}$ &	27$\pm$4 & S &  \\
23 & & 09:03:53.598	& 0.10  &	6.8$\times10^{-6}$	& 2.2$\times10^{-7}$  & 18$\pm$3 & S &  \\
24 & & 09:03:54.878	& 0.29   &	3.1$\times10^{-6}$	& 3.8$\times10^{-7}$  & 33$\pm$3 & S  &  \\
25 & & 09:04:00.270	& 0.22   &	8.2$\times10^{-6}$	& 6.0$\times10^{-7}$ &	25$\pm$2 & S &  \\
26 & & 09:04:10.446	& 0.50  &	$\ge2.6\times10^{-6}$ & 5.0$\times10^{-7}$  & 31$\pm$3 & S &  \\
27 & & 09:04:33.038	& 0.70  &	$\ge3.8\times10^{-5}$ & 1.0$\times10^{-5}$ &	27$\pm$1 & S &  \\
28 & & 09:05:14.318	& 0.19  &	$\ge1.4\times10^{-6}$ & 1.4$\times10^{-7}$  & 25$\pm$6 & S &  \\
29 & & 09:05:54.382	& $\le0.75$& $\ge4.2\times10^{-7}$ & 2.1$\times10^{-7}$  & 30$\pm$7 & S &  \\
30 & & 09:06:31.502	& $\le0.5$ & $\ge3\times10^{-6}$ & 1.2$\times10^{-6}$ &	27$\pm$1 & S &  \\
31 & & 09:06:44.558	& $\le0.75$ & $\ge1.4\times10^{-6}$ & 3.8$\times10^{-7}$ &	16$\pm$4 & S &  \\
32& & 09:08:51.3 & $\le4$	  & $\ge1\times10^{-6}$ & 3.5$\times10^{-6}$ &	25$\pm$1 & S &  \\
33 & 980530b  & 11:28:52.737	& 0.28  &	1.4$\times10^{-5}$	& 1.5$\times10^{-6}$ &	18$\pm1$ & T  &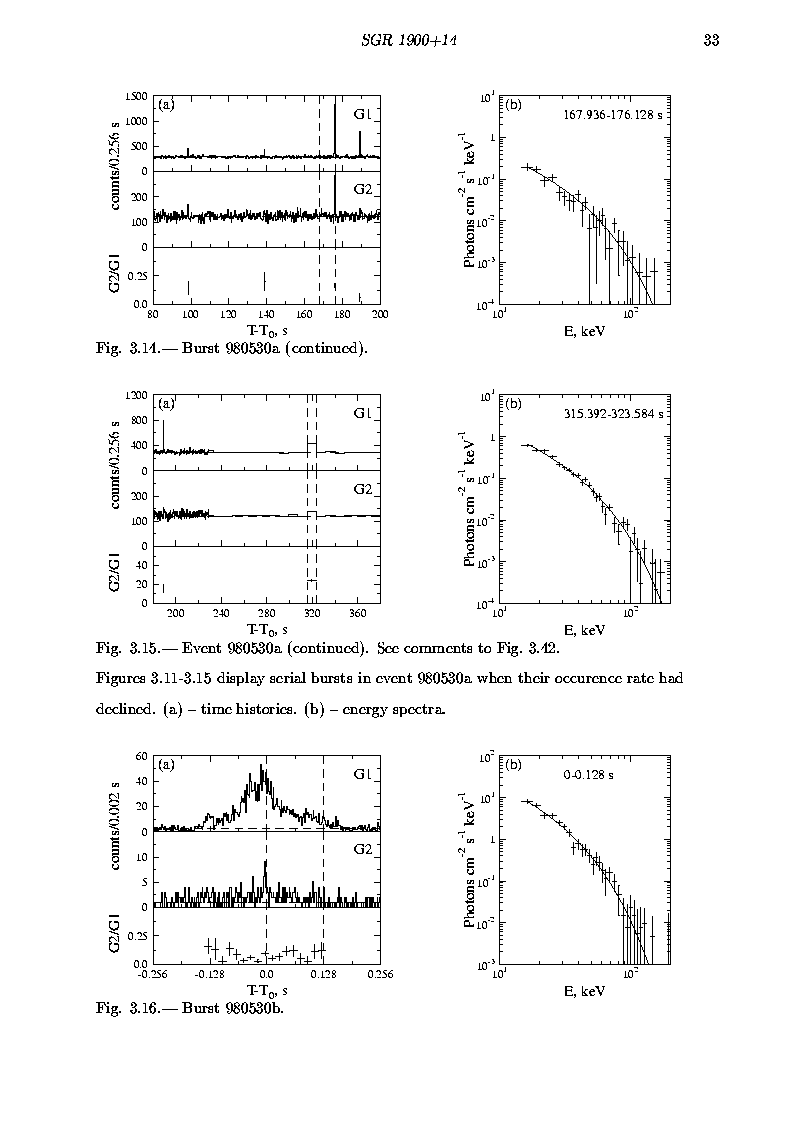  \\
34 & 980530c  & 12:47:18.778	& 0.37  &	2.3$\times10^{-5}$ & 2.8$\times10^{-6}$ &	27$\pm1$ & T &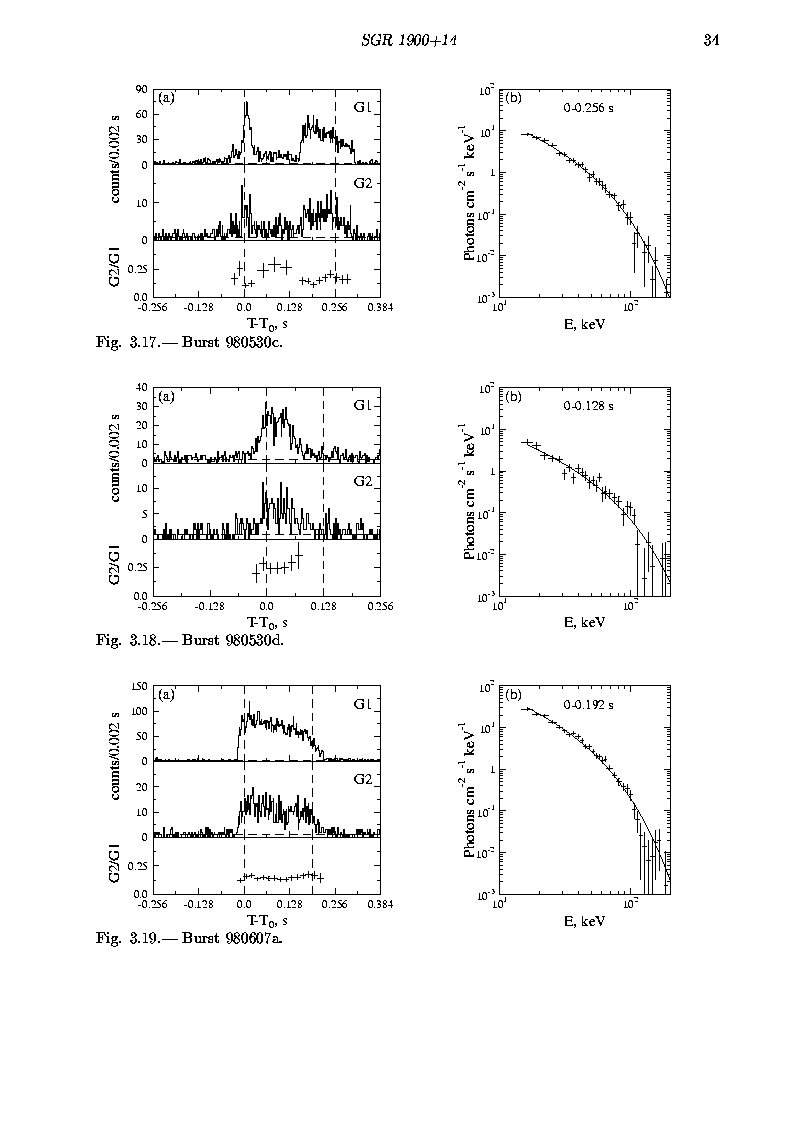  \\
35 & 980530d  & 23:27:37.482	& 0.18  &	9.3$\times10^{-6}$	& 7.9$\times10^{-7}$ &	36$\pm3$ & T & SGRfig14.png \\
36 & 980601  & 13:33:55 & $\le3$	   & \nodata	               & 3.3$\times10^{-7}$ &	$\le35$ & B & \nodata\\
37 & 980604  & 03:13:43 & $\le3$	   & \nodata	               & 2.3$\times10^{-7}$ &	$\le100$ & B & \nodata \\
38 & 980607a  & 08:32:32.512	& 0.26  & 3.4$\times10^{-5}$ & 5.4$\times10^{-6}$ &	25$\pm1$ & T & SGRfig14.png \\
39 & 980607b  & 21:14:15.040 & 0.10  & 2.1$\times10^{-5}$ & 4.3$\times10^{-7}$ & 29$\pm$2 & T  & 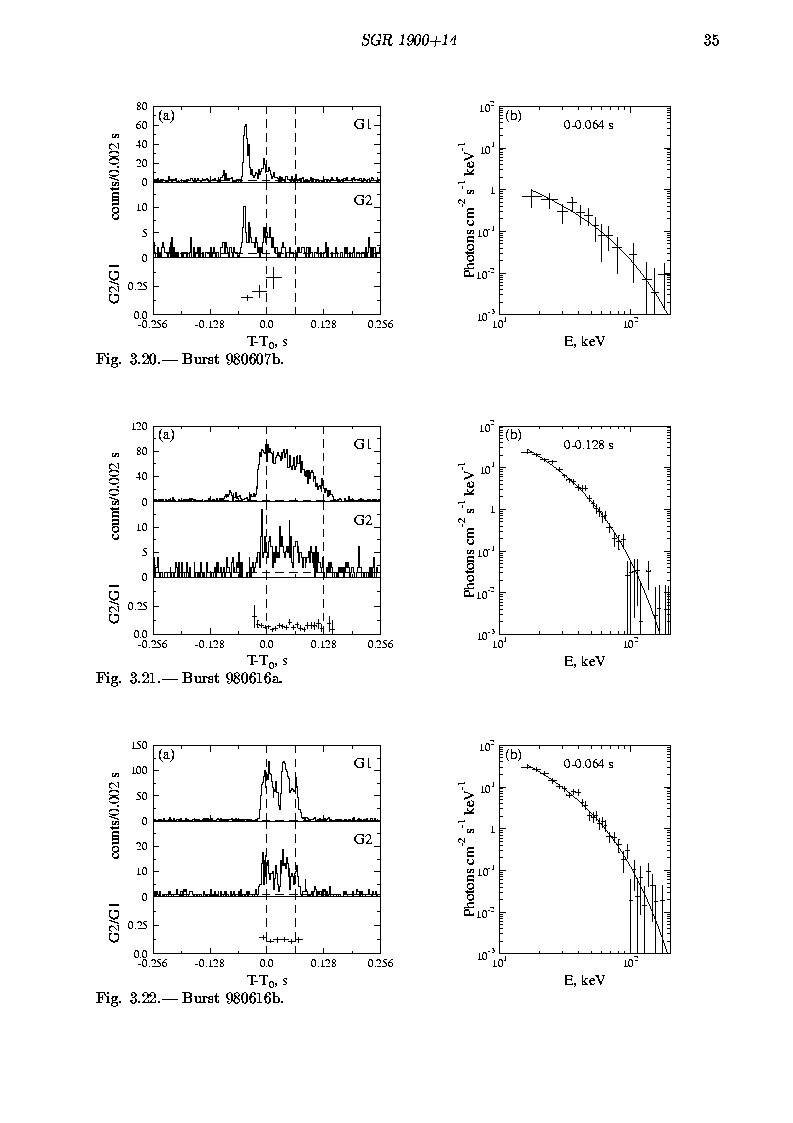 \\
40 & 980616a  & 01:50:26.582	& 0.25  & 2.7$\times10^{-5}$ & 3.1$\times10^{-6}$ &19$\pm1$ & T  & SGRfig15.png \\
41 & 980616b  & 17:22:59.574	& 0.11 &	4.0$\times10^{-5}$ & 2.3$\times10^{-6}$ &	22$\pm1$ & T  & SGRfig15.png \\
42 & 980719a  & 08:26:24.995	& 0.04 &	1.1$\times10^{-5}$ & 2.9$\times10^{-7}$ &	32$\pm$3 & T & 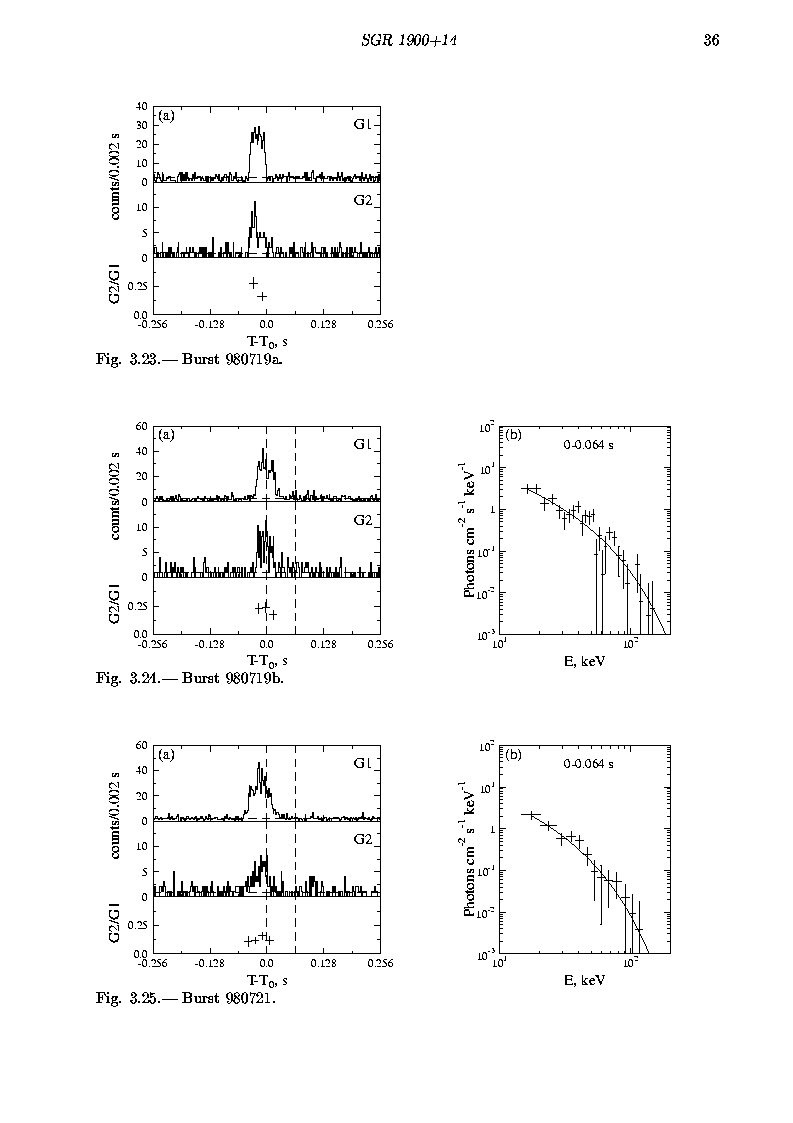 \\
43 & 980719b  & 16:40:11.217	& 0.05 &	1.1$\times10^{-5}$ & 4.2$\times10^{-7}$ & 31$\pm$2& T  & SGRfig16.png \\
44 & 980721  & 03:19:00.412	& 0.07  &	1.2$\times10^{-5}$ & 5.1$\times10^{-7}$ & 24$\pm$2 & T & SGRfig16.png \\
45 & 980821  & 21:30:24.989	& 0.19   &	2.4$\times10^{-5}$ & 3.0$\times10^{-6}$	 & 28$\pm 1$ & T & 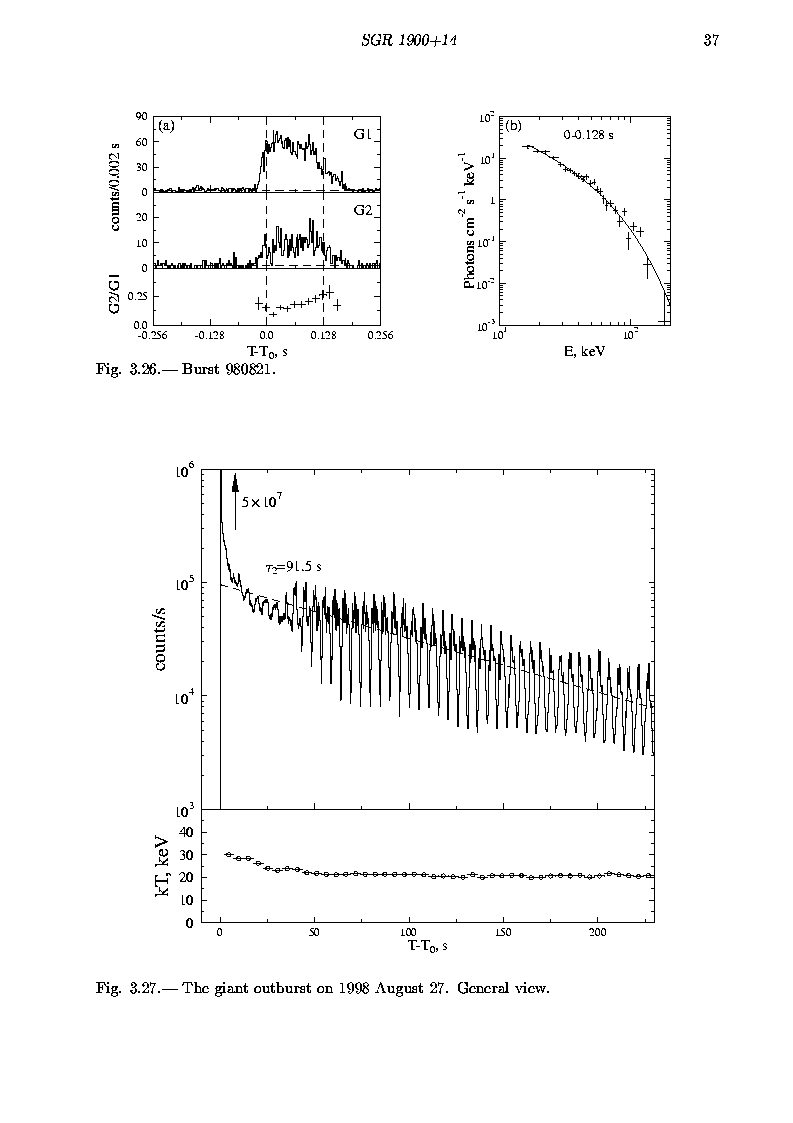 \\
46 & 980827  & 10:22:17.578	& & & $>9.7\times10^{-3}$ & & T & SGRfig17.png \\
  & & & & & &  & & --SGRfig22.png\\
 & Initial pulse &  & 0.35 & $>3.1\times 10^{-2}$ & $>5.5\times 10^{-3}$ & 300$\to$20 &  &  \\ 
 & Pulsating tail & & $>230$ &  & 4.2$\times 10^{-3}$ & $\sim20$ &  &  \\ 
47 & 980828  & 23:14:10.680	& 0.10 &	3.1$\times10^{-5}$ & 1.4$\times10^{-6}$ & 20$\pm 1$ & T & 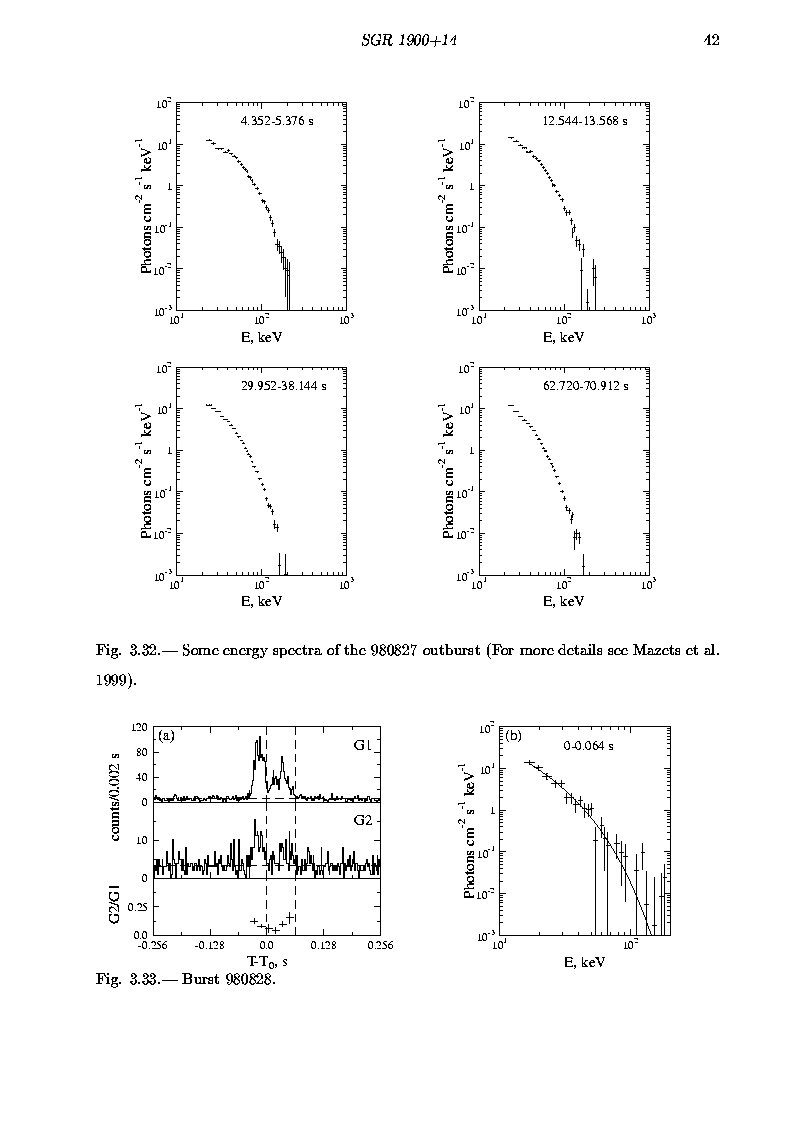 \\
48 & 980829a  & 10:16:34.068	& 3.50 &	2.4$\times10^{-5}$ & 3.6$\times10^{-5}$	& 20$\pm 1$ & T & 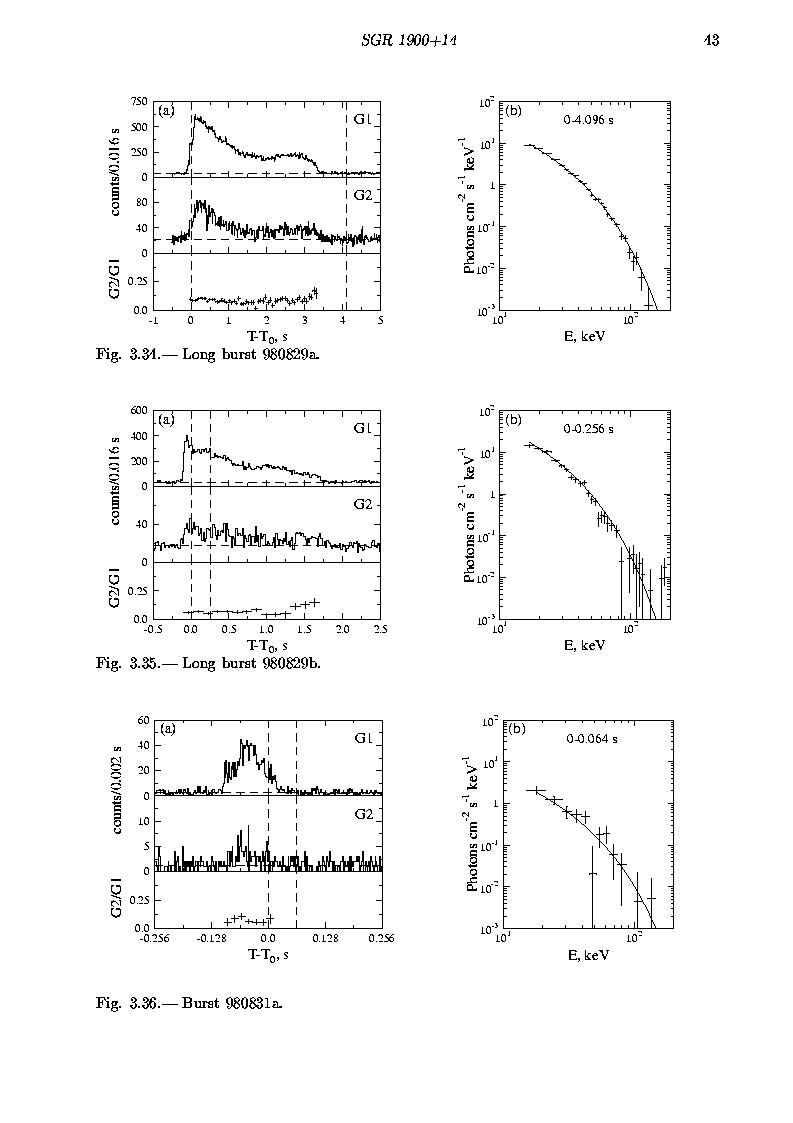 \\
49 & 980829b  & 18:18:45.166	& 1.85 &	1.4$\times10^{-5}$ & 1.1$\times10^{-5}$	& 17$\pm 1$ & T  & SGRfig23.png \\
50 & 980831a  & 15:47:38.156	& 0.12 &	1.5$\times10^{-5}$ & 9.6$\times10^{-7}$ & 18$\pm1$& T & SGRfig23.png \\
51 & 980831b  & 19:13:47.249	& 0.19  &	1.3$\times10^{-5}$ & 1.4$\times10^{-6}$	& 18$\pm 1$ & TS & 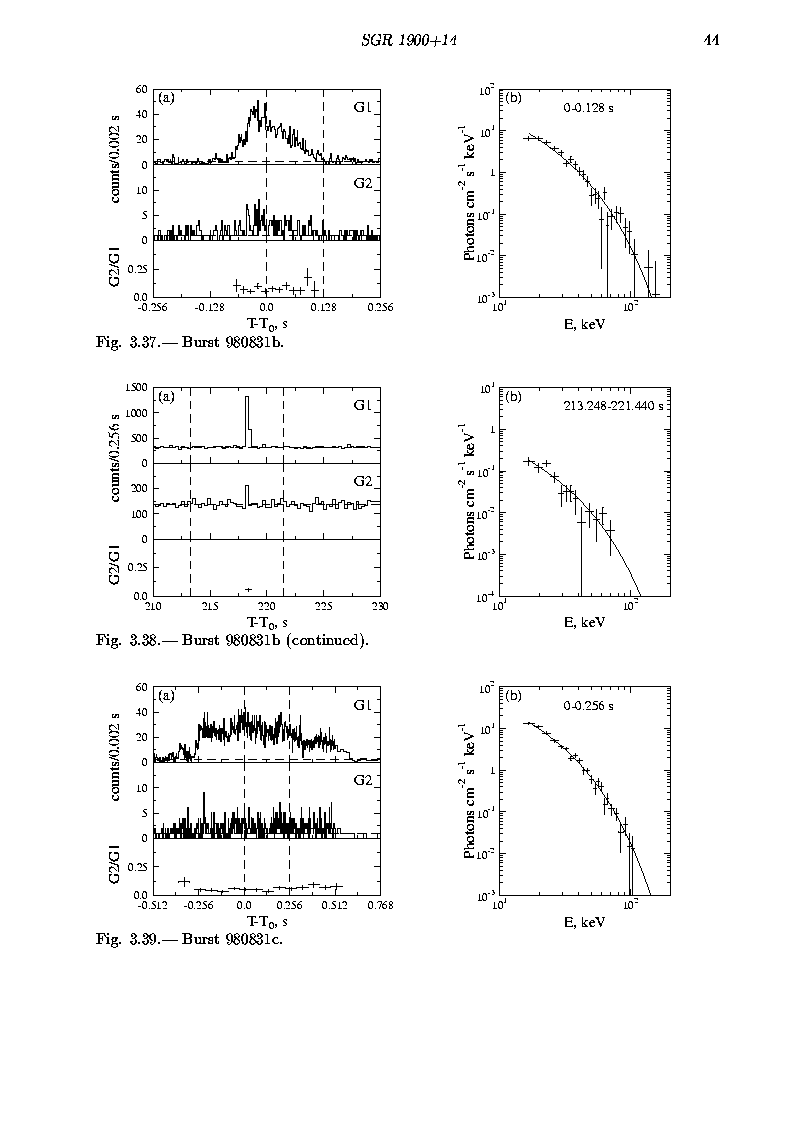 \\
52 &          & 19:17:25.361 & $\le0.5$	& $\ge3\times10^{-6}$ & 1.0$\times10^{-6}$	& 17$\pm$2 & S &  \\
53 & 980831c  & 22:22:06.721	& 1.00  & 1.1$\times10^{-5}$	& 6.0$\times10^{-6}$	& 17$\pm 1$ & T & SGRfig24.png \\
54 & 980901a  & 04:54:16.387	& 0.53  &	1.7$\times10^{-5}$	& 2.6$\times10^{-6}$ &	17$\pm1$ & T & 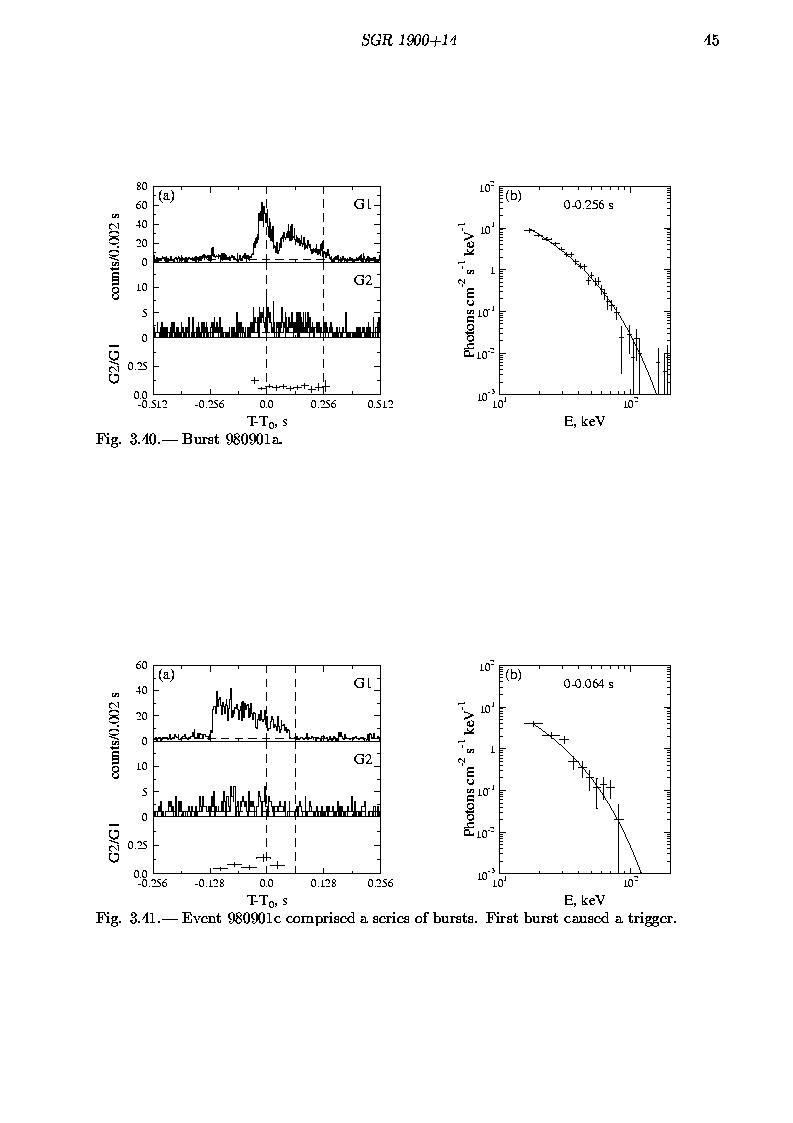 \\
55 & 980901b & 16:13:02 & \nodata	   & \nodata	              & 2.6$\times10^{-7}$	 & $\le25$ & B & \nodata\\
56 & 980901c  & 17:00:32.594 & 0.18  &	9.5$\times10^{-6}$	& 1.1$\times10^{-6}$     & 18$\pm1$ & TS & SGRfig25.png \\
 & & & & & &  & & --SGRfig27.png \\
57 &   & 17:03:49.202 & $\le0.5$   & $\ge4\times10^{-6}$  & 1$\times10^{-6}$	 & 17$\pm$1 & S & \\
58 &  & 17:06:25.1 & $\sim32$	   & $\ge10^{-5}$         & 2$\times10^{-5}$	 & $22\pm1$& S & \\
59 & 980902a  & 17:20:19.649	& 0.35  &	1.1$\times10^{-5}$	& 2.4$\times10^{-6}$	 & 20$\pm 1$ & T & 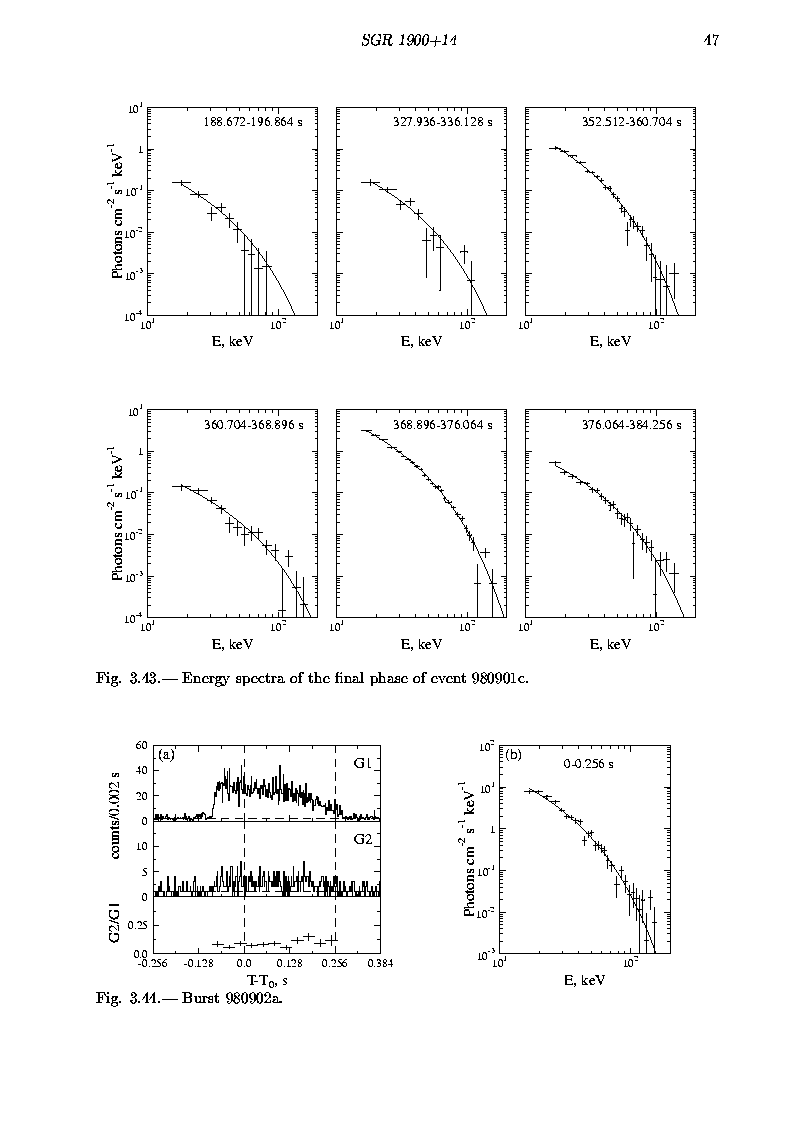 \\
60 & 980902b  & 19:47:31.640 & 0.25 &	1.7$\times10^{-5}$	& 2.1$\times10^{-6}$	 & 20$\pm 1$ & T & 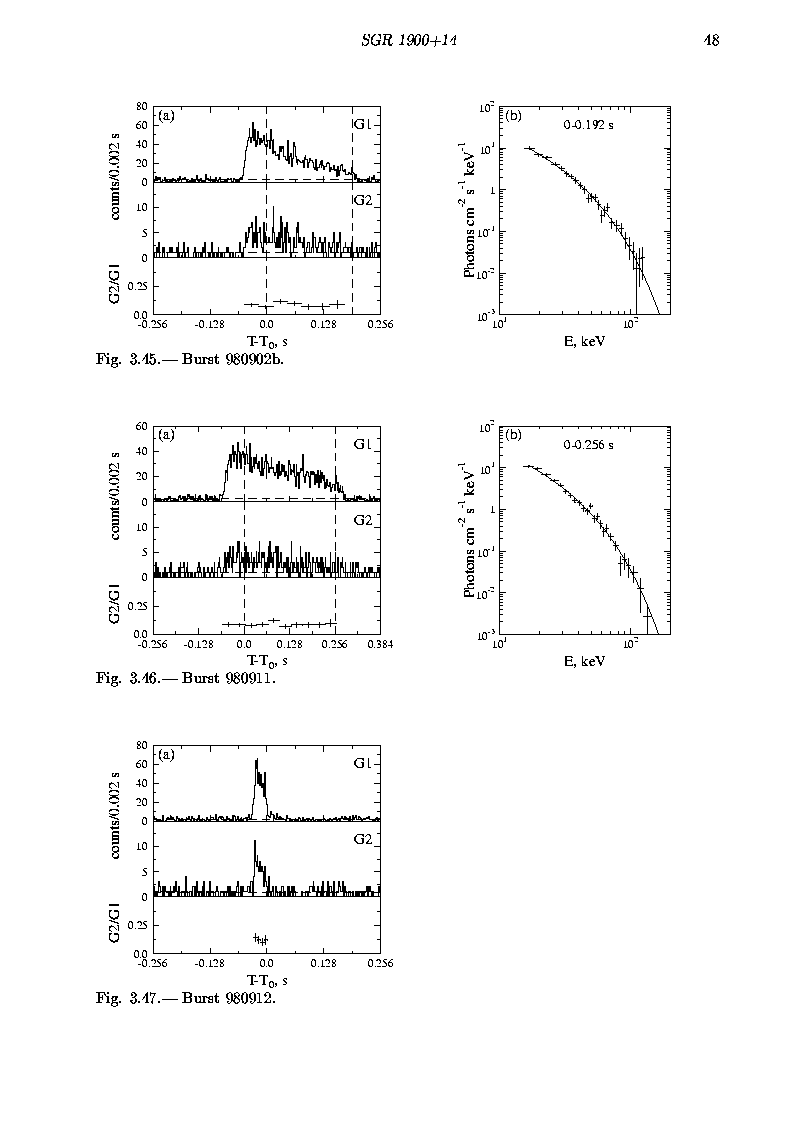 \\
61 & 980908  & 06:43:41 & $\le3$	   & \nodata	              & 1.8$\times10^{-7}$	 & $\le35$ & B & \nodata \\
62 & 980911  & 18:22:38.838	& 0.34 &	1.5$\times10^{-5}$	& 2.6$\times10^{-6}$	 & 20$\pm 1$ & T & SGRfig28.png \\
63 & 980912  & 21:17:19.646	& 0.05 &	2.1$\times10^{-5}$	& 4.7$\times10^{-7}$	 & 23$\pm$2 & T  & SGRfig28.png \\
64 & 980915a & 18:17:42.117	& 0.25 &	3.2$\times10^{-5}$ & 4.0$\times10^{-6}$	 & 18$\pm1$ & T & 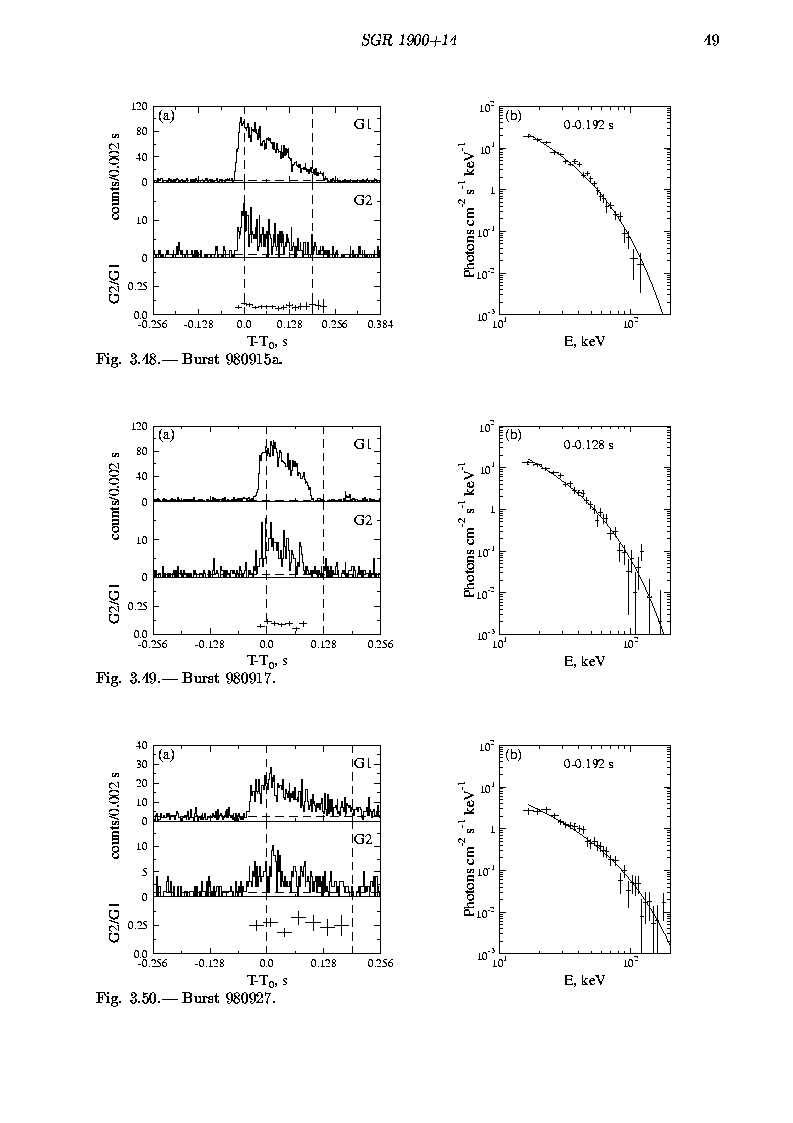 \\
65 & 980915b & 21:40:27 & $\le3$	   & \nodata                 & 3.7$\times10^{-7}$	 & $\le30$ & B & \nodata \\
66 & 980917  & 09:45:41.987	& 0.22  & 2.6$\times10^{-5}$ & 2.5$\times10^{-6}$    & 20$\pm	1$ & T & SGRfig29.png \\
67 & 980925  & 12:03:47 & $\le3$	   & \nodata	              & 1.3$\times10^{-7}$ & $\le40$ & B & \nodata \\
68 & 980927  & 02:07:55.646	& 0.23  &	8.3$\times10^{-6}$ & 1.0$\times10^{-6}$   & 35$\pm$2 & T & SGRfig29.png \\
69 & 981013a  & 00:31:50.530 & 0.20 &	1.8$\times10^{-5}$ & 1.9$\times10^{-6}$	 & 18$\pm 1$ & TS & 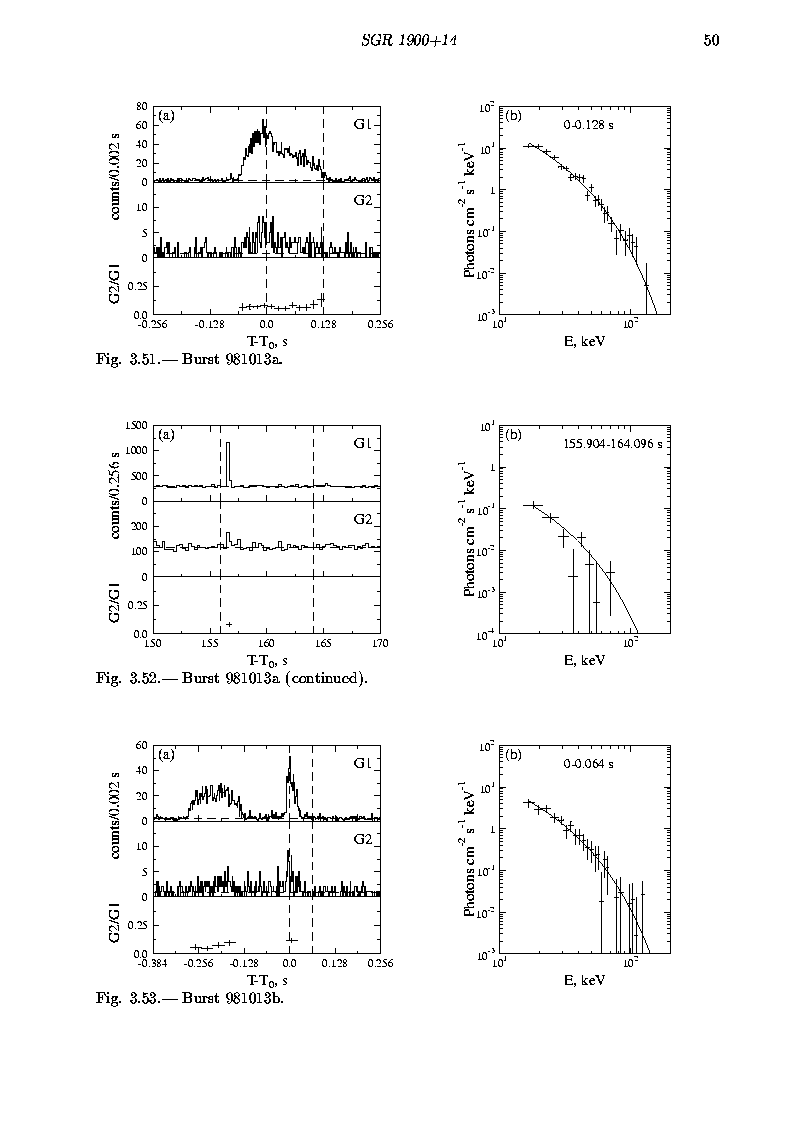 \\
70 &  & 00:31:57.830 & 0.06	   & 1.8$\times10^{-6}$     & 1.0$\times10^{-7}$     & 35$\pm$6 & S & \nodata \\
71 &  & 00:34:26.946 & $\le0.5$   & $\ge2.3\times10^{-6}$ & 7.0$\times10^{-7}$	 & 19$\pm$2 & S &  SGRfig30.png\\
72 &  & 00:35:39.906 & $\le0.3$   & $\ge1\times10^{-6}$ & 2$\times10^{-7}$	 & 20$\pm$4 & S & \nodata\\
73 & 981013b  & 13:57:47.699	& 0.31 &	1.4$\times10^{-5}$	& 1.3$\times10^{-6}$	 & 20$\pm 1$ & T & SGRfig30.png \\
74 & 981022  & 15:40:46.627	& 0.40 &	2.1$\times10^{-6}$	& 3.9$\times10^{-7}$	 & 110$\pm$30 & T & 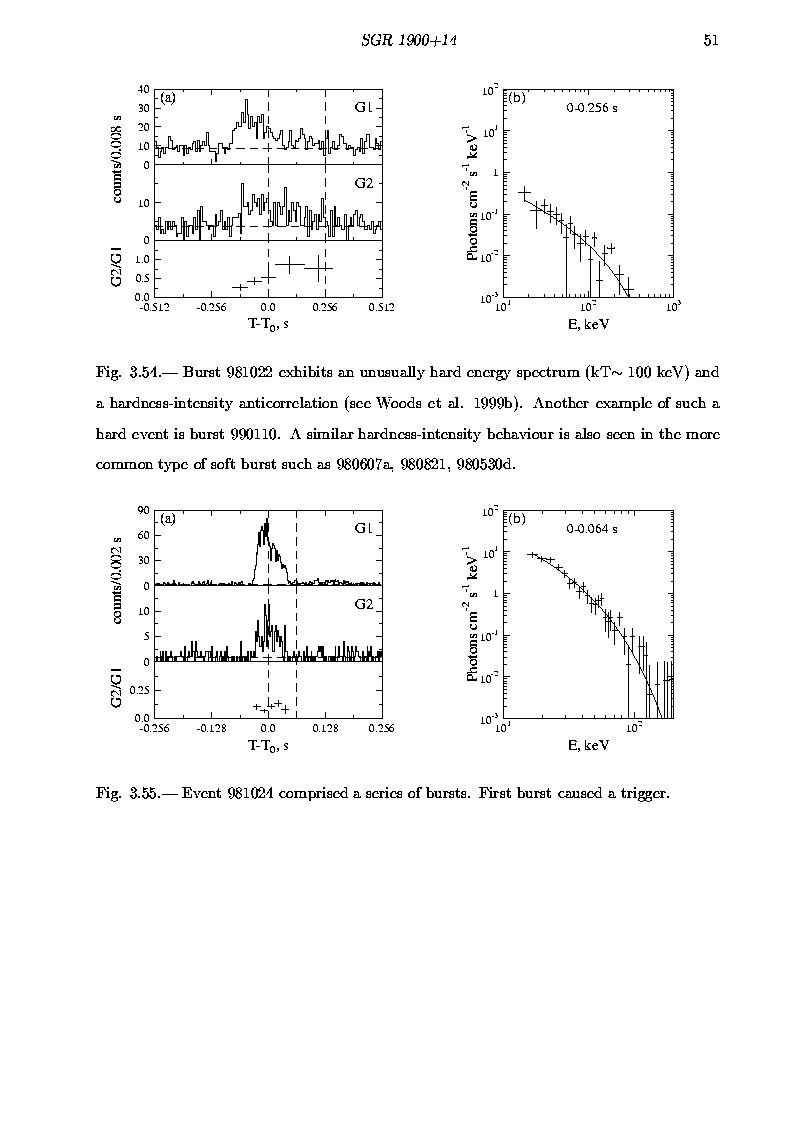 \\
75 & 981024  & 01:22:01.645	& 0.18 &	2.4$\times10^{-5}$	& 1.1$\times10^{-6}$	 & 20$\pm 1$ & TS & SGRfig31.png \\
  & & & & & & & & --SGRfig33.png \\
76 & & 01:25:32.077	& $\le0.25$   & $\ge5\times10^{-7}$   & 1.5$\times10^{-7}$	 & 24$\pm$8 & S &  \\
77 & & 01:25:43.597& $\le0.5$   & $\ge7\times10^{-7}$   & 3.1$\times10^{-7}$	 & 24$\pm$4 & S & \\
78 & & 01:27:21.4  	& $\le16$   & \nodata                 & 1.0$\times10^{-6}$	 & 32$\pm$6   & S & \\
79 & & 01:29:07.9  	& $\le24$   & \nodata                 & 1.2$\times10^{-5}$	 & 25$\pm$1 & S &  \\
80 & & 01:29:48.9  	& $\le16$   & \nodata                 & 1.6$\times10^{-6}$	 & 26$\pm$2   & S & \\
81 & 981028a  & 20:20:53.276	& 0.11 &	1.8$\times10^{-5}$	& 8.1$\times10^{-7}$	 & 23$\pm$2 & T & 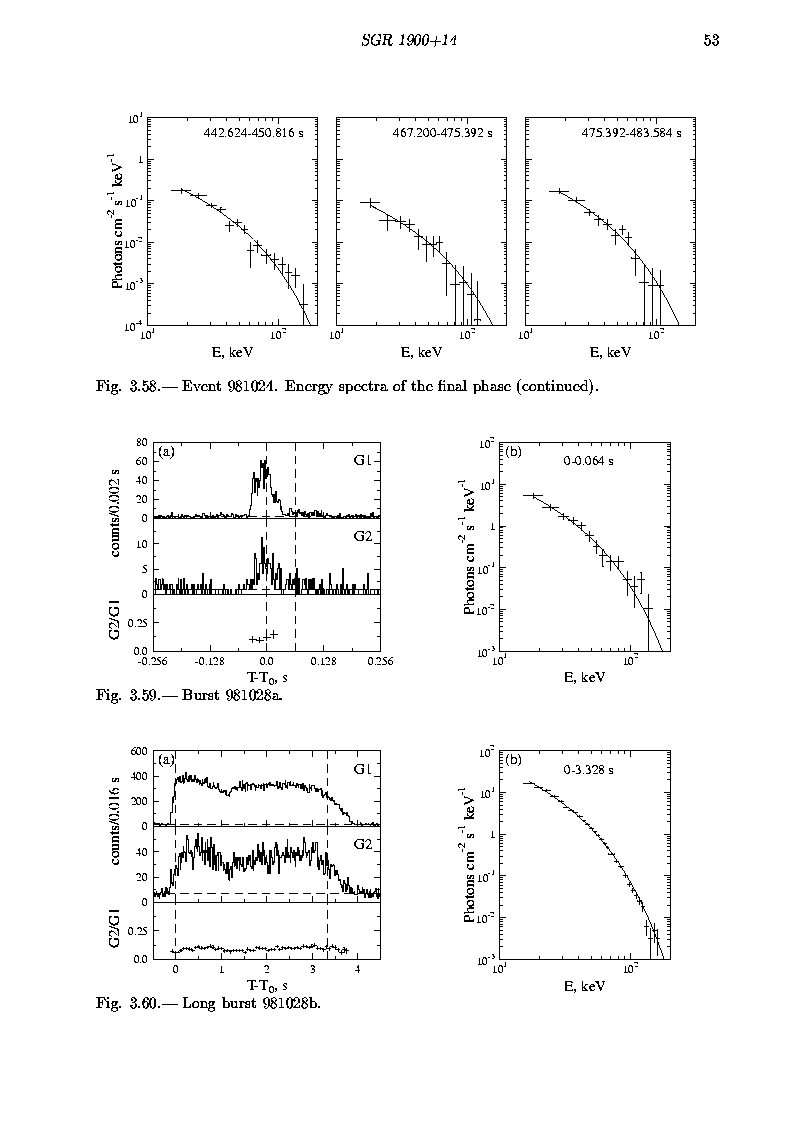 \\
82 & 981028b  & 23:03:39.997	& 4.0 &	1.6$\times10^{-5}$	& 4.8$\times10^{-5}$	      & 22$\pm 1$ & T & SGRfig33.png \\
83 & 981029  & 21:36:19.861	& 0.13 &	2.4$\times10^{-5}$	& 1.8$\times10^{-6}$	 & 22$\pm 1$ & T & 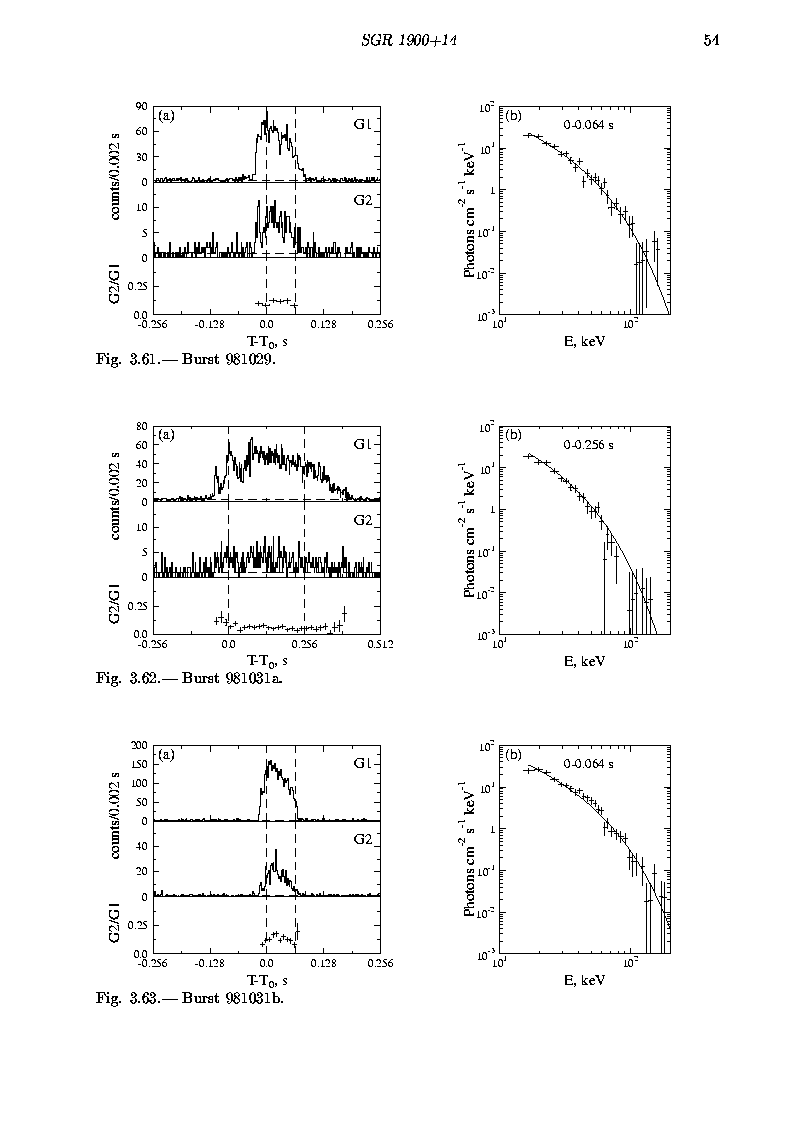 \\
84 & 981031a  & 18:32:08.605	& 0.45  &	1.7$\times10^{-5}$	& 5.1$\times10^{-6}$	 & 17$\pm 1$ & T & SGRfig34.png \\
85 & 981031b  & 22:14:07.013	& 0.09  &	5.3$\times10^{-5}$	& 3.2$\times10^{-6}$	 & 24$\pm 1$ & T & SGRfig34.png \\
86 & 981109  & 21:19:09.763	& 0.07 &	2.4$\times10^{-5}$	& 1.0$\times10^{-6}$	      & 20$\pm	1$ & T & 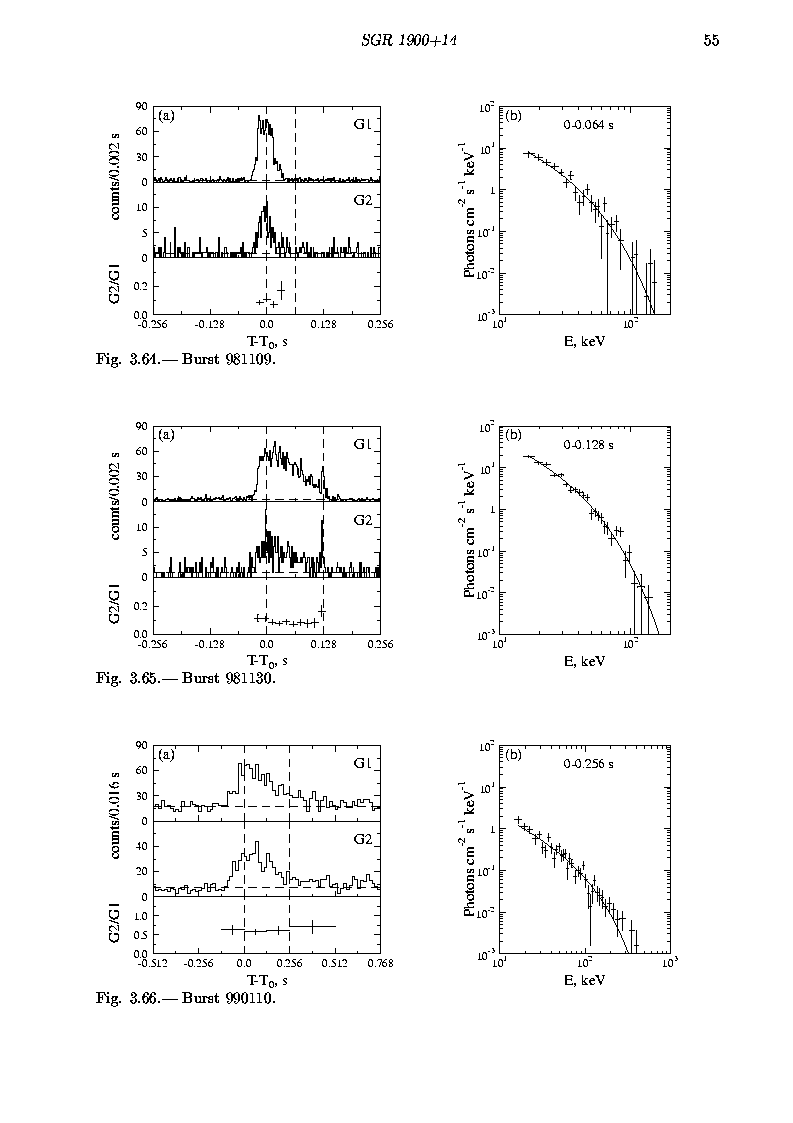 \\
87 & 981130  & 03:28:53.044	& 0.17 &	2.1$\times10^{-5}$	& 2.1$\times10^{-6}$	 & 20$\pm 1$ & T & SGRfig35.png \\
88 & 990110  & 08:39:01.078	& 0.48 &	3.0$\times10^{-6}$	& 8.0$\times10^{-7}$	      & 75$\pm	6$ & T & SGRfig35.png \\
\enddata
\tablenotetext{a}{It can be seen from this Table that SGR 1900+14 emitted
bursts of three types: single bursts, 
series of bursts, and giant outburst (August 27 event). 
These types of events are displayed below in
separate sets of Figures.}

\end{deluxetable}

\setcounter{subsection}{4}
\setcounter{table}{0}
\begin{deluxetable}{lllcccccc}
\tabletypesize{\scriptsize}
\tablecaption{SGR 1806-20}
\tablehead{
\colhead{N} & \colhead{Burst}& \colhead{T$_0$}& \colhead{$\Delta$T} & \colhead{P$_{max}$} &%
\colhead{S} & \colhead{kT} & \colhead{Comments}& \colhead{Figures}  \\
& \colhead{name} &\colhead{h:m:s UT}& \colhead{(s)}&\colhead{(erg cm$^{-2}$ s$^{-1}$)} &%
\colhead{(erg cm$^{-2}$)} & \colhead{(keV)} &&\colhead{PNG file}
}
\startdata
1 & 790107  & 05:31:42.015 & 0.25 & 6.7$\times10^{-6}$ & 1.4$\times10^{-6}$ & 25$\pm$2 & T & 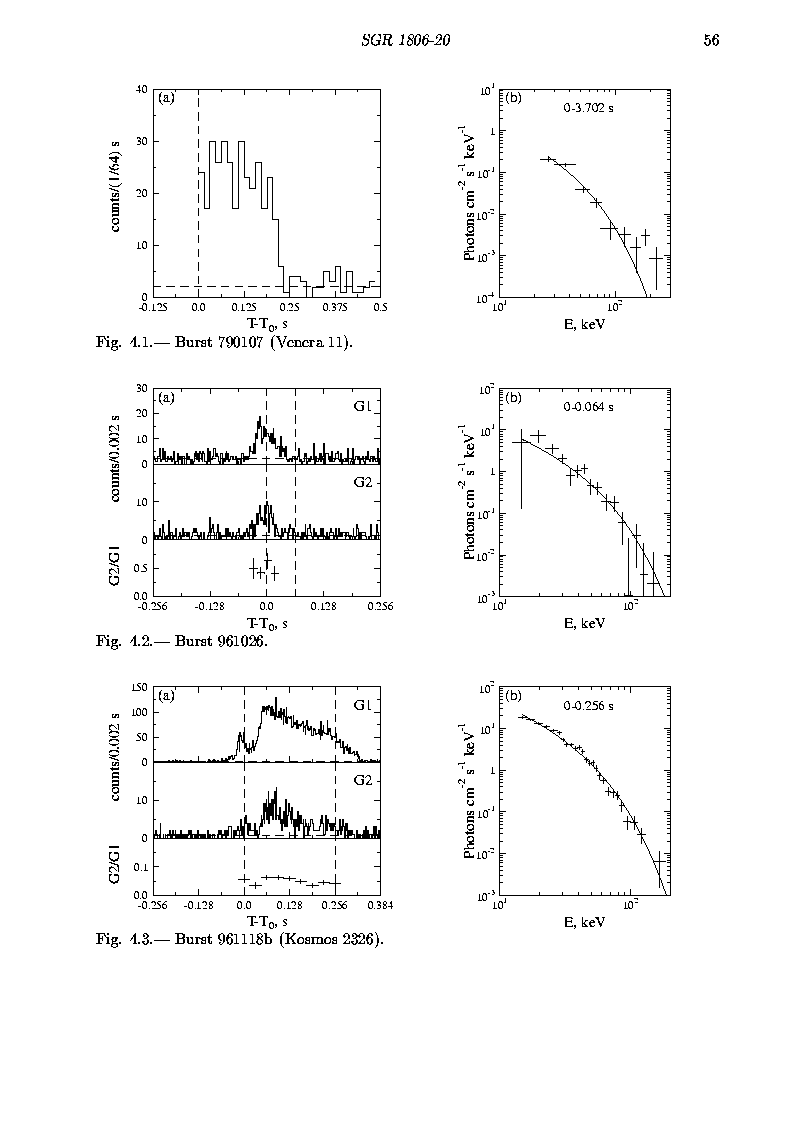 \\
2 & 961026  & 01:33:51.940 & 0.06 & 1.8$\times10^{-5}$ & 8.0$\times10^{-7}$ & 24$\pm$2 &T & SGRfig36.png \\
3 & 961105  & 17:02:18     &  $\le1.5$  & \nodata & 7$\times10^{-7}$ & 21$\pm$5 & B & \nodata \\
4 & 961118a  & 04:52:35 & $\le1.5$ & \nodata & 4$\times10^{-7}$ & $\le30$ & B& \nodata \\
5 & 961118b  & 05:25:10.181 & 0.37 & 3.8$\times10^{-5}$ & 5.8$\times10^{-6}$ & 20$\pm$1 & TS & SGRfig36.png \\
6 & & 05:25:56.133 & $\sim$0.10 & $\ge4.0\times10^{-6}$ & 4.5$\times10^{-7}$ & 26$\pm$4 & S&  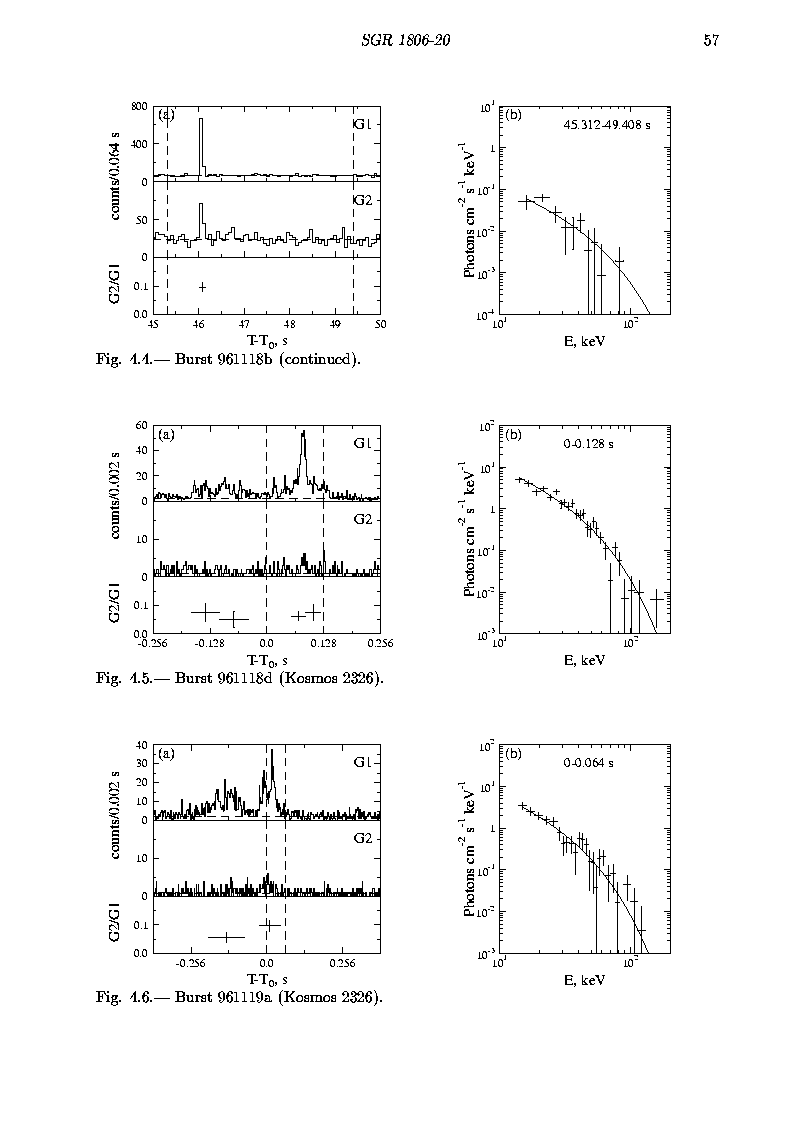\\
7 & 961118c  & 11:26:12 & $\le1.5$ & \nodata & 3$\times10^{-7}$ & $\le30$ & B & \nodata \\
8 & 961118d  & 16:04:57.507 & 0.31 & 1.8$\times10^{-5}$ & 1.0$\times10^{-6}$ & 20$\pm$2 & T& SGRfig37.png \\
9 & 961119a  & 03:04:08.608 & 0.24 & 8$\times10^{-6}$ & 6$\times10^{-7}$ & 24$\pm$3 & T& SGRfig37.png \\
10 & 961119b  & 05:30:25.967 & 0.83 & 2.0$\times10^{-5}$ & 1.0$\times10^{-5}$ & 18$\pm$1 & T& 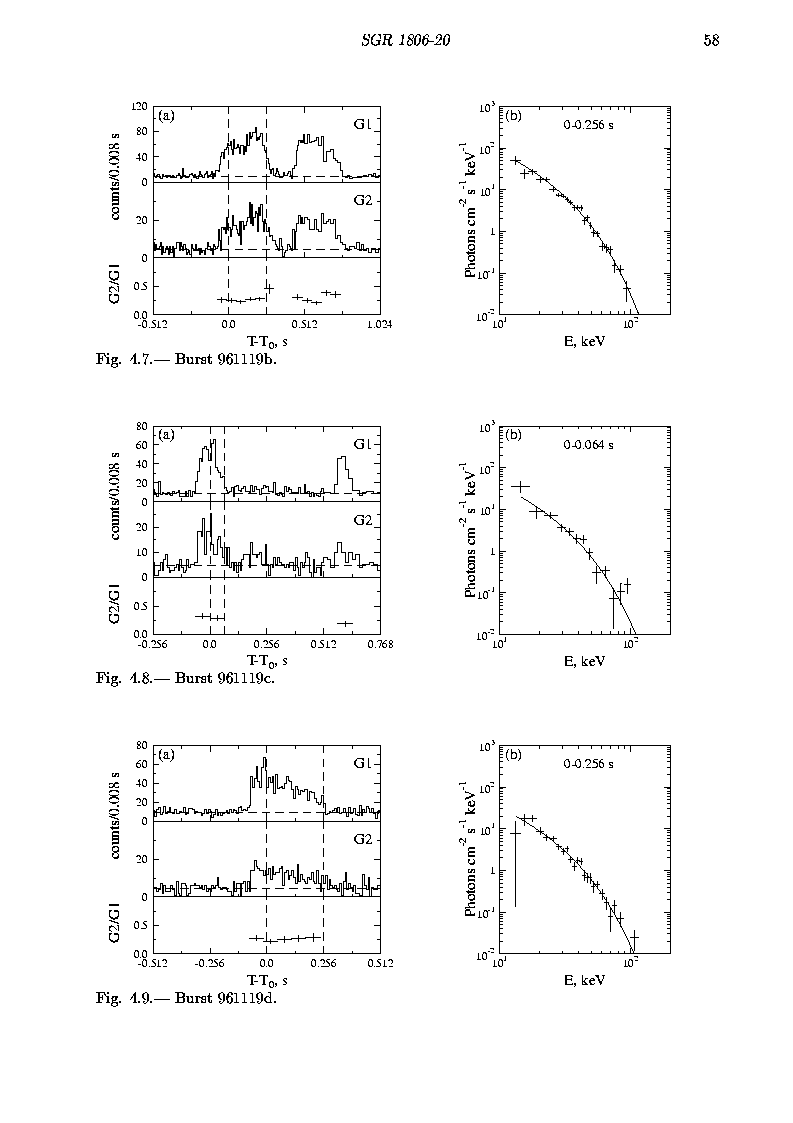 \\
11 & 961119c  & 06:22:07.737 & 0.30 & 1.3$\times10^{-5}$ & 2.2$\times10^{-6}$ & 20$\pm$1 & T & SGRfig38.png \\
12 & 961119d  & 07:24:04.153 & 0.32 & 1.2$\times10^{-5}$ & 2.6$\times10^{-6}$ & 17$\pm$2 & T& SGRfig38.png \\
13 & 961121  & 07:05:39 & $\le1.5$ & \nodata & 2$\times10^{-7}$ & $\le40$ & B& \nodata \\
14 & 961122  & 02:17:38.595 & 0.16 & 2.7$\times10^{-5}$ & 2.6$\times10^{-6}$ & 21$\pm$2 & T& 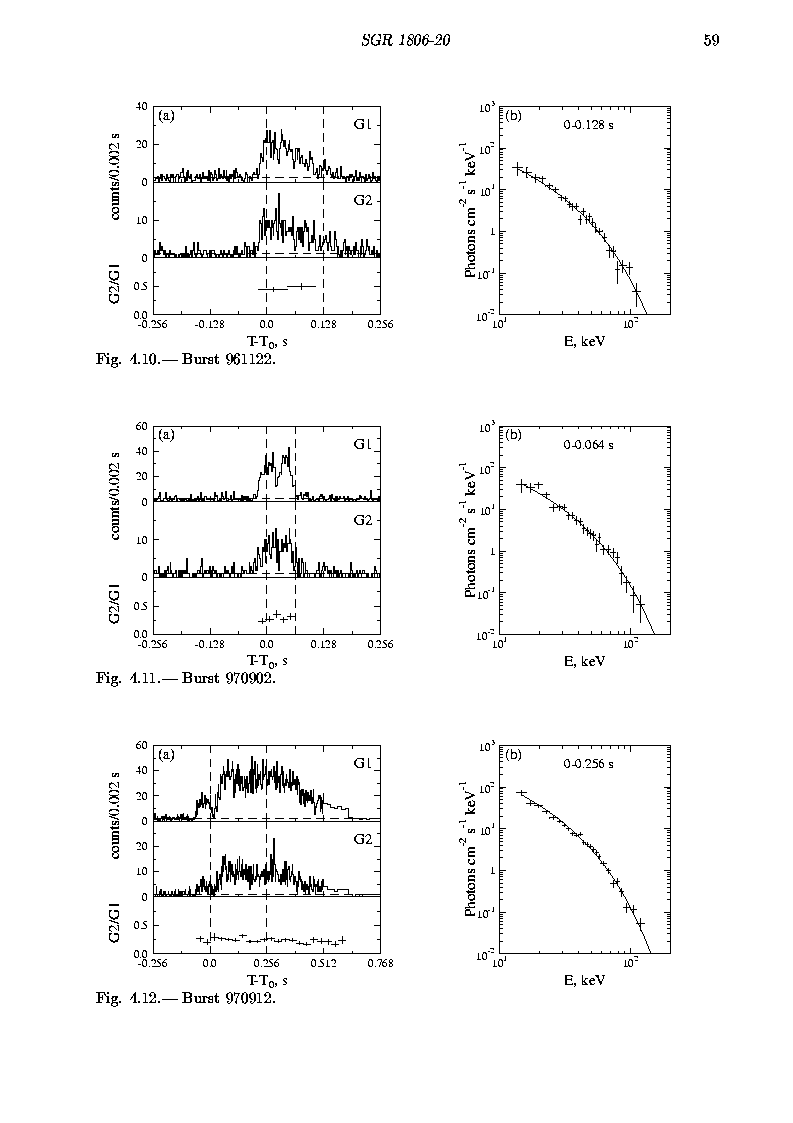 \\
15 & 970124  & 07:47:21 & $\le3$ & \nodata & 6$\times10^{-7}$ & $\le30$ & B&  \nodata\\
16 & 970902  & 10:55:46.779 & 0.09 & 3.2$\times10^{-5}$ & 2.0$\times10^{-6}$ & 22$\pm$1 & T & SGRfig39.png \\
17 & 970912  & 06:09:24.080 & 0.68 & 3.6$\times10^{-5}$ & 1.5$\times10^{-5}$ & 20$\pm$1 & T & SGRfig39.png \\
18 & 981017  & 15:13:43.353 & 0.20 & 2.5$\times10^{-5}$ & 3.0$\times10^{-6}$ & 22$\pm$2 & T & 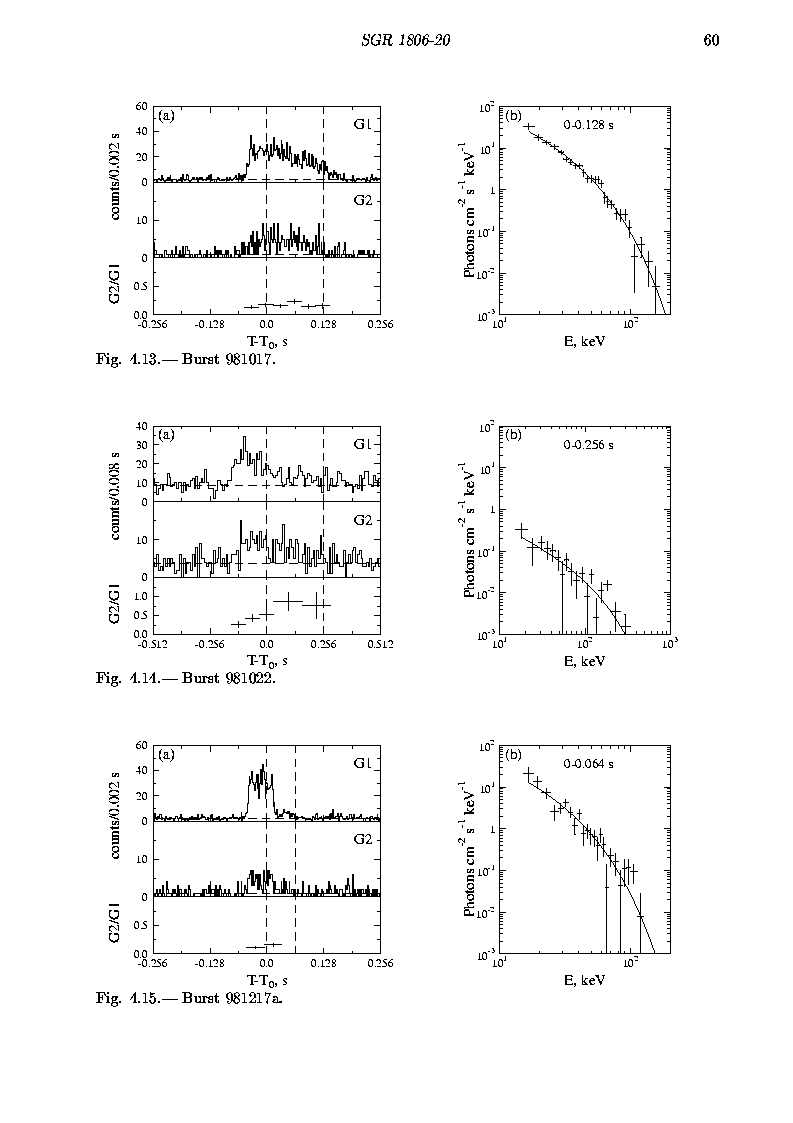 \\
19 & 981022  & 05:21:53.701 & 0.06 & 2.5$\times10^{-5}$ & 9.0$\times10^{-7}$ & 23$\pm$2 & T & SGRfig40.png \\
20 & 981217a  & 13:21:18.615 & 0.10 & 3.0$\times10^{-5}$ & 1.7$\times10^{-6}$ & 19$\pm$2 & T& SGRfig40.png \\
21 & 981217b  & 19:13:58.678 & 0.08 & 5.9$\times10^{-5}$ & 2.4$\times10^{-6}$ & 20$\pm$2 & T& 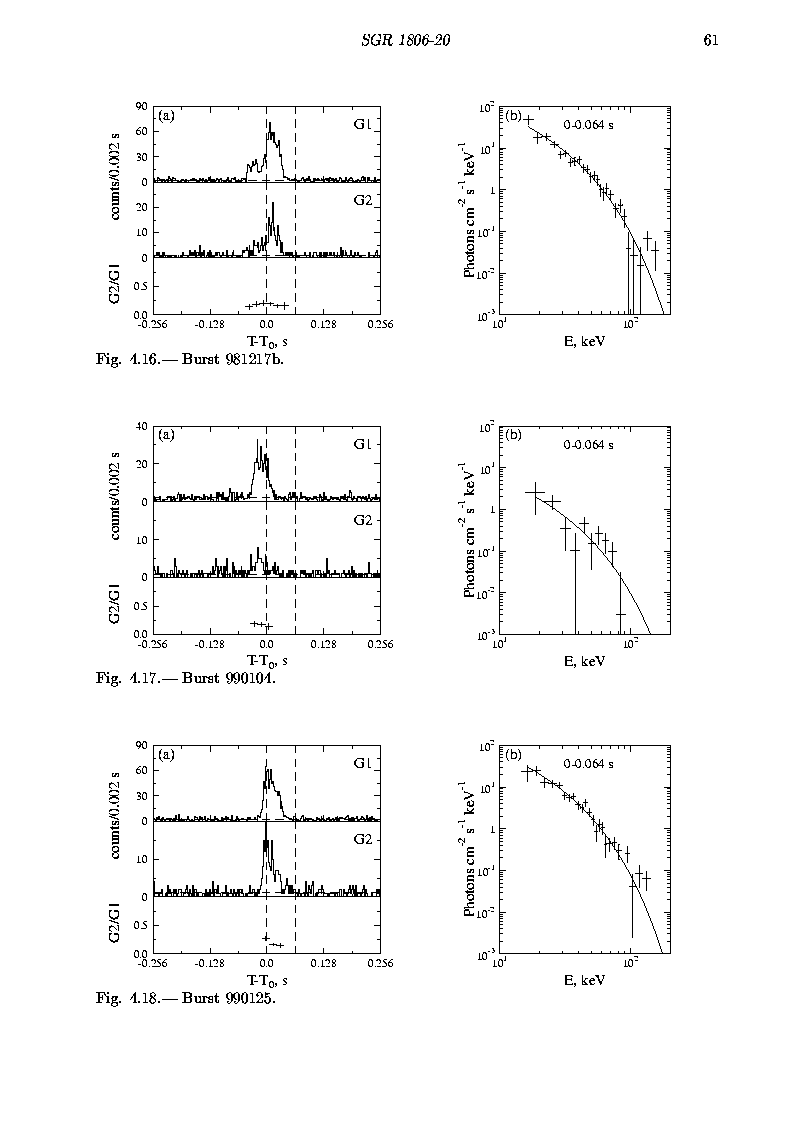 \\
22 & 990104  & 01:50:25.592 & 0.05 & 2$\times10^{-5}$ & 7$\times10^{-7}$ & 22$\pm$3 & T& SGRfig41.png \\
23 & 990125  & 08:33:56.811 & 0.05 & 6.1$\times10^{-5}$ & 1.9$\times10^{-6}$ & 20$\pm$2 & T& SGRfig41.png \\
24 & 990205  & 08:44:11 & $\le3$ & \nodata & 8$\times10^{-7}$ & $\le30$ & B& \nodata \\
25 & 990206  & 02:10:32.228 & 0.06 & 4.5$\times10^{-5}$ & 1.4$\times10^{-6}$ & 20$\pm$2 & T& 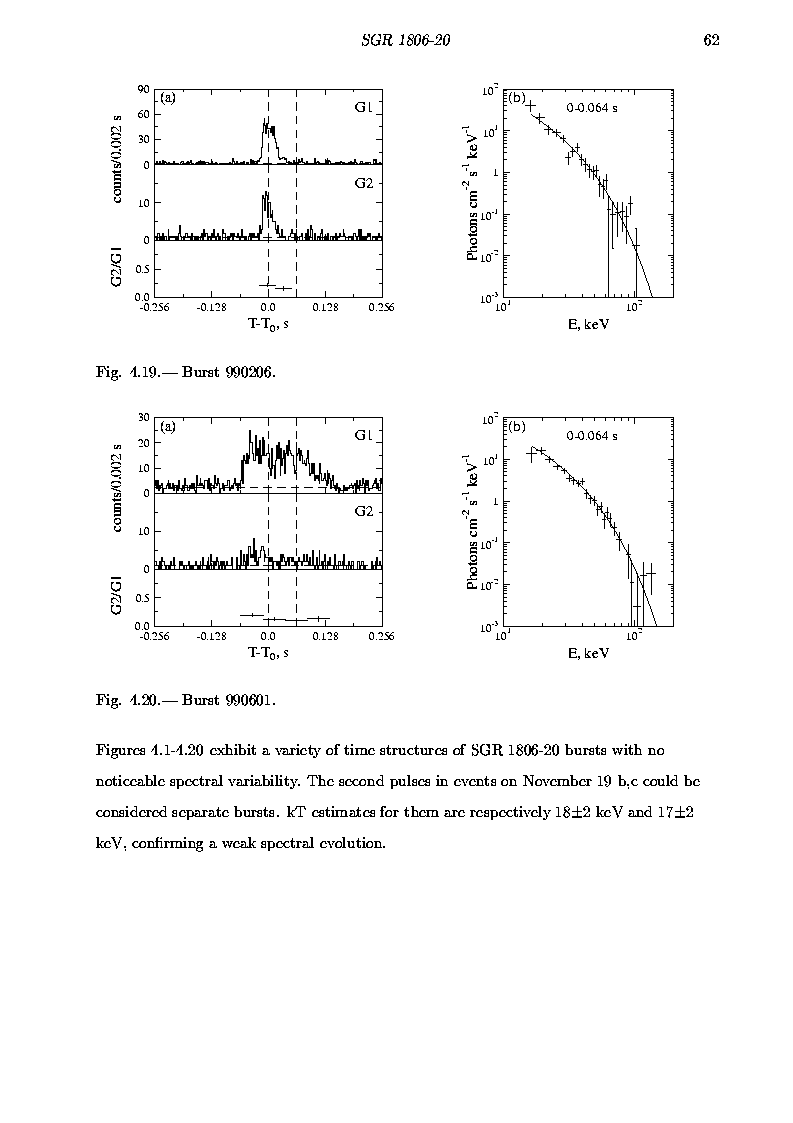 \\
26 & 990601  & 10:00:08.743 & 0.17 & 1.5$\times10^{-5}$ & 1.9$\times10^{-6}$ & 19$\pm$2 & T& SGRfig42.png \\
\enddata
\end{deluxetable}

\setcounter{subsection}{5}
\setcounter{table}{0}
\begin{deluxetable}{lllccccccc}
\tabletypesize{\scriptsize}
\tablecaption{SGR 1627-41\tablenotemark{a}}
\tablehead{
\colhead{N} & \colhead{Burst}& \colhead{T$_0$}& \colhead{$\Delta$T} & \colhead{P$_{max}$} &%
\colhead{S} & \colhead{kT} & \colhead{Comments}& \colhead{Figures} \\
& \colhead{name} &\colhead{h:m:s UT}& \colhead{(s)}&\colhead{(erg cm$^{-2}$ s$^{-1}$)} &%
\colhead{(erg cm$^{-2}$)} & \colhead{(keV)}   & &\colhead{PNG file}
&  
}
\startdata
1 & 980617a  & 18:53:29 & $\le3$ & \nodata & 5$\times10^{-7}$ & 19$\pm$5 & B& \nodata\\
2 & 980617b  & 18:58:14 & $\le3$ & \nodata & 4$\times10^{-7}$ & $\le20$ & B& \nodata \\
3 & 980617c  & 19:04:02 & $\le3$ & \nodata & 1$\times10^{-6}$ & 19$\pm$2 & B& \nodata \\
4 & 980617d  & 19:58:31.969 & 0.15 & 5.1$\times10^{-5}$ & 4.6$\times10^{-6}$ &29$\pm$2 & TS& 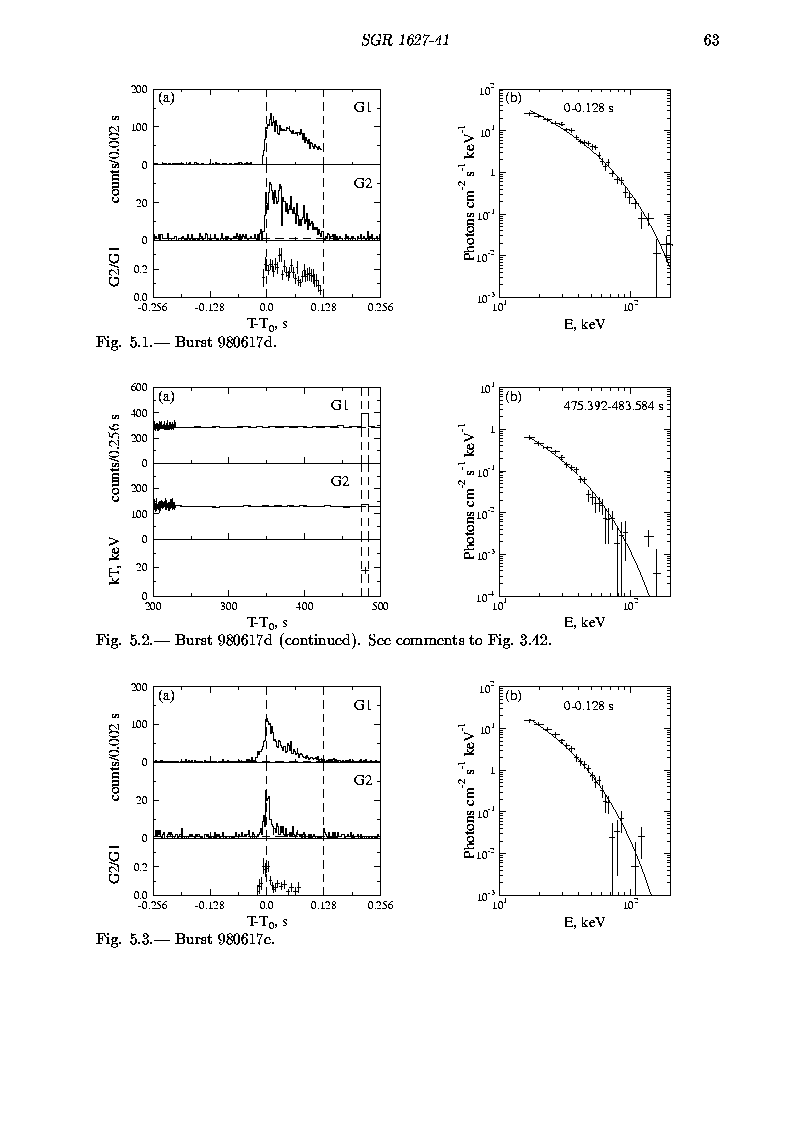 \\
5 &          & 20:06:27.4 & $\le4$ & \nodata & 3$\times10^{-6}$ & 18$\pm$2 & S& SGRfig43.png \\
6 & 980617e  & 21:04:40.001 & 0.15 & 4.6$\times10^{-5}$ & 2.1$\times10^{-6}$ &20$\pm$1 & TS& SGRfig43.png \\
7 &          & 21:07:24.353 & $\sim1$ & $\ge6 \times 10^{-6}$ & 2$\times10^{-6}$ & 21$\pm$1 & S& 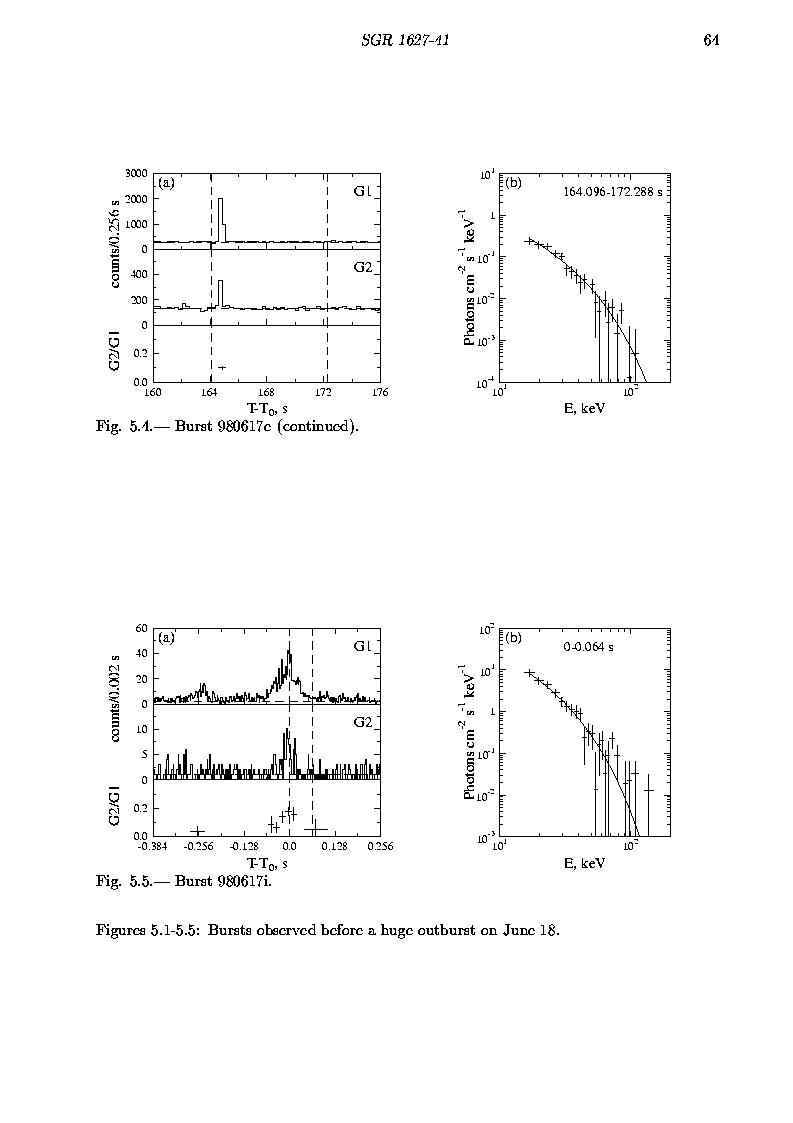 \\
8 & 980617f  & 21:17:06 & $\le4$ & \nodata & 4$\times10^{-6}$ &\nodata & H& \nodata \\
9 & 980617g  & 21:37:21 & $\le4$ & \nodata & 5$\times10^{-6}$ &\nodata & H&  \nodata\\
10 & 980617h  & 21:57:10 & $\le4$ & \nodata & 4$\times10^{-6}$ &\nodata & H&  \nodata\\
11 & 980617i  & 22:54:08.930 & 0.30 & 1.6$\times10^{-5}$ & 1.1$\times10^{-6}$ & 21$\pm$2 & T& SGRfig44.png \\
12 & 980618a  & 00:06:11 & $\le3$ & \nodata & 3$\times10^{-7}$ & $\le30$ & B&  \nodata\\
13 & 980618b  & 00:15:35 & $\le3$ & \nodata & 5$\times10^{-7}$ & $\le20$ & B&  \nodata\\
14 & 980618c  & 00:51:04 & $\le3$ & \nodata & 2$\times10^{-7}$ & $\le70$ & B&  \nodata\\
15 & 980618d  & 01:42:33.495 & 0.50 & 0.028 & 7.6$\times10^{-4}$ & 20-120 & T& 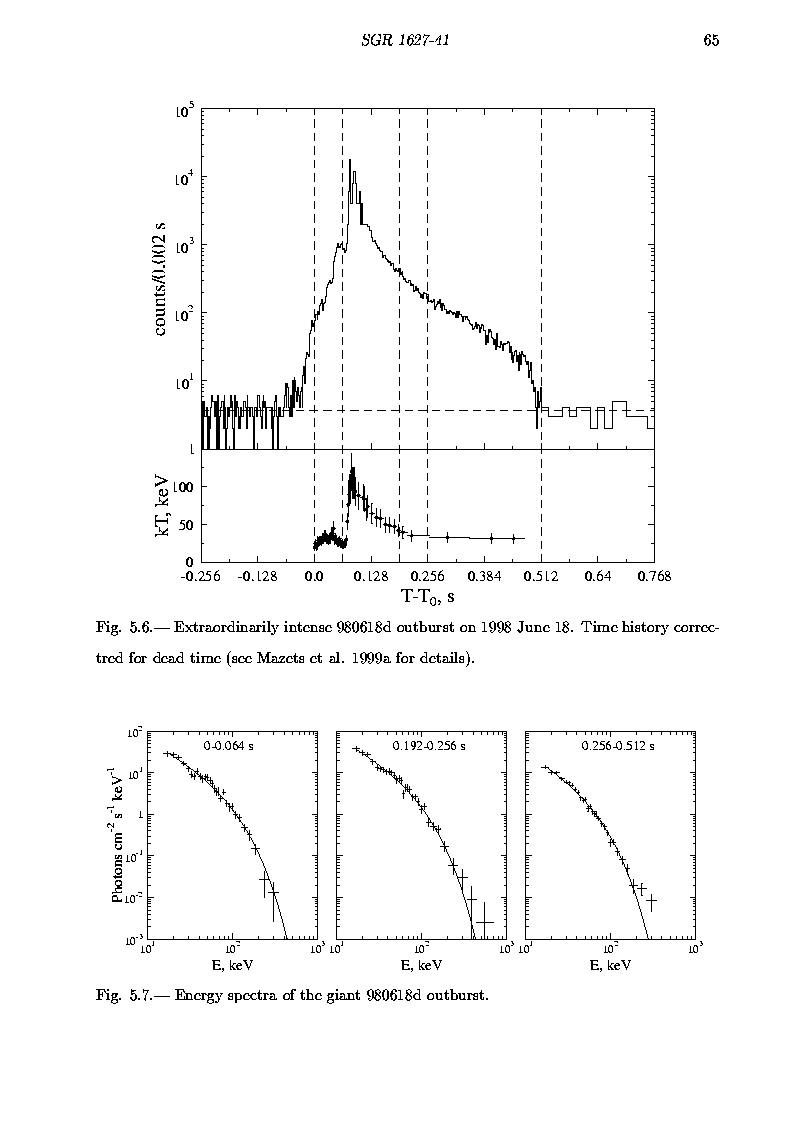 \\
16 & 980618e  & 03:07:49 & \nodata & \nodata & 5$\times10^{-7}$ & $\le30$ & B& \nodata \\
17 & 980618f  & 03:25:16.450 & 0.22 & 1.5$\times10^{-5}$ & 1.0$\times10^{-6}$ & 22$\pm$2 & TS& 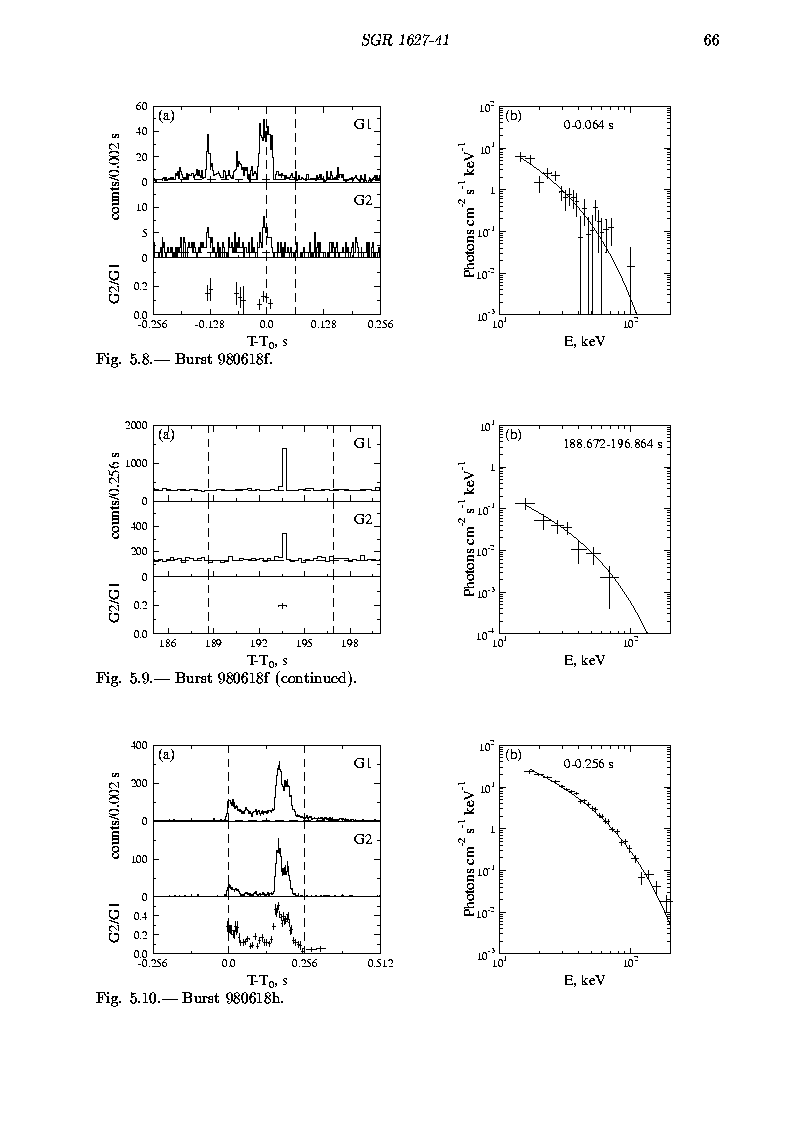 \\
18 &   & 03:28:30.010 & $\le0.3$ & $\ge4 \times 10^{-6}$ & 1$\times10^{-6}$ &  30$\pm$2 & S& SGRfig46.png \\
19 & 980618g & 04:04:24 & $\le4$ & \nodata & $\ge5\times10^{-5}$ &\nodata & H& \nodata\\
20 & 980618h  & 04:30:29.019 & 0.47 & 1.5$\times10^{-4}$ & 1.1$\times10^{-5}$ & 38$\pm$1 & T& SGRfig46.png \\
21 & 980618i  & 04:34:20 & $\le3$ & \nodata & 4$\times10^{-7}$ &\nodata & B& \nodata \\
22 & 980618j  & 05:42:09 & $\le3$ & \nodata & 5$\times10^{-7}$ & 38$\pm$8 & B& \nodata \\
23 & 980618k  & 06:16:33.125 & 0.07 & 1.8$\times10^{-5}$ & 6.3$\times10^{-7}$ & 26$\pm$2 & T& 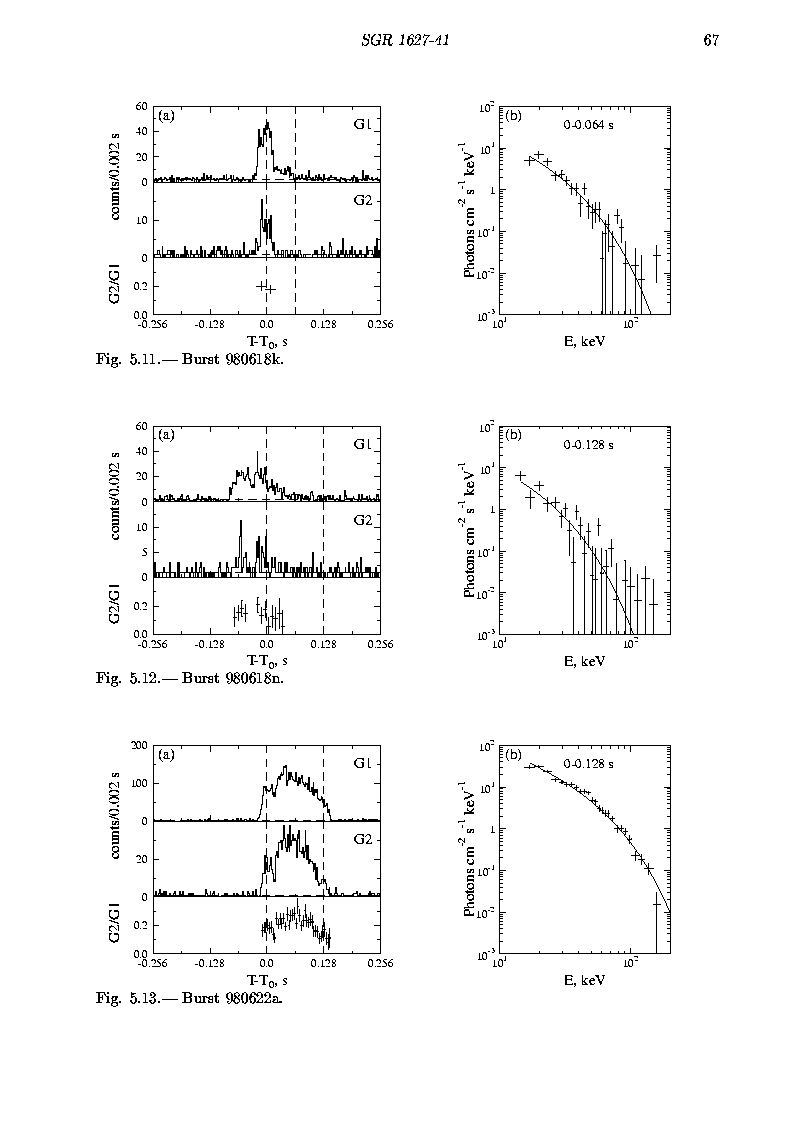 \\
24 & 980618l  & 07:34:08 & $\le3$ & \nodata & 3$\times10^{-7}$ & $\le25$ & B& \nodata \\
25 & 980618m  & 12:19:23 & $\le3$ & \nodata & 2$\times10^{-7}$ & $\le40$ & B& \nodata \\
26 & 980618n  & 16:38:04.665 & 0.15 & 8.7$\times10^{-6}$ & 7.7$\times10^{-7}$ & 23$\pm$2 & T& SGRfig47.png \\
27 & 980618o  & 17:51:16 & $\le3$ & \nodata & 7$\times10^{-7}$ & 27$\pm$4 & B& \nodata \\
28 & 980622a  & 13:29:56.270 &  0.16 & 6.3$\times10^{-5}$ & 6.0$\times10^{-6}$ & 31$\pm$1 & T& SGRfig47.png \\
29 & 980622b & 14:11:25 & $\le4$ & \nodata & 7$\times10^{-7}$ &\nodata & H&  \nodata\\
30 & 980622c  & 18:56:39.058 & 0.10 & 4.0$\times10^{-5}$ & 1.1$\times10^{-6}$ & 29$\pm$2 & T& 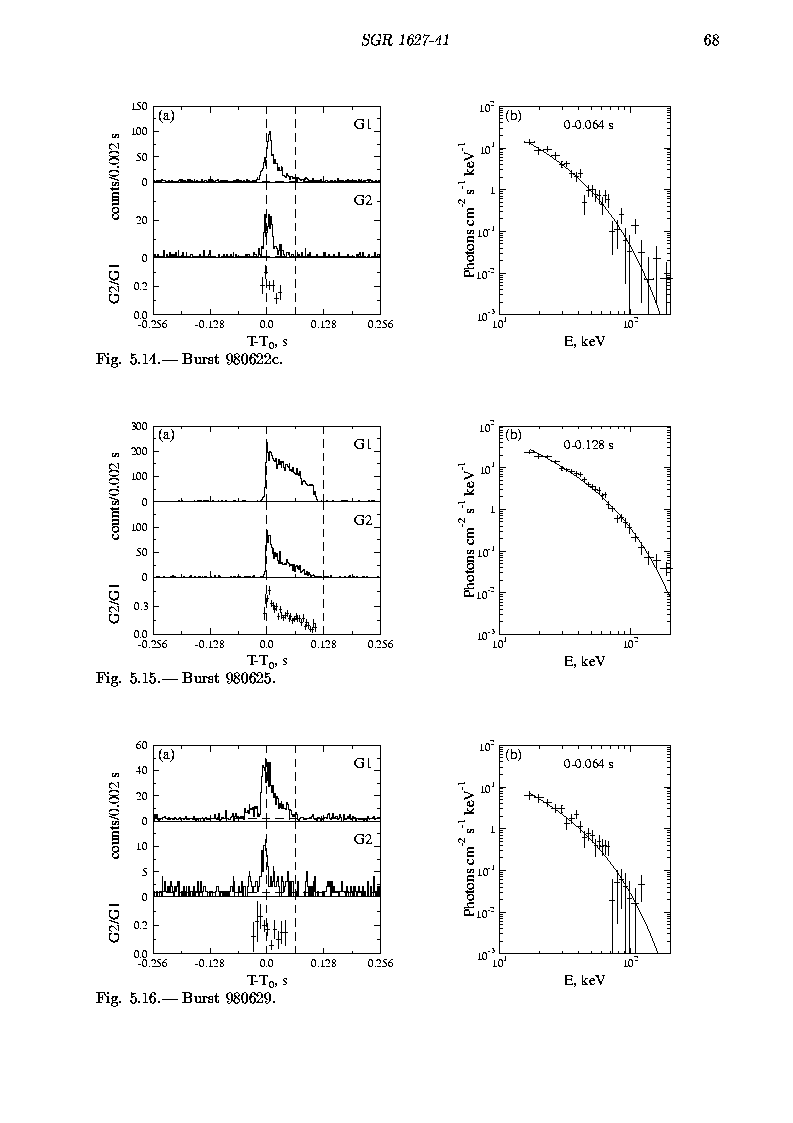 \\
31 & 980625  & 10:56:19.575 & 0.13 & 1.1$\times10^{-4}$ & 6.3$\times10^{-6}$ & 33$\pm$1 & T& SGRfig48.png \\
32 & 980629  & 07:25:06.756 & 0.10 & 1.8$\times10^{-5}$ & 7.0$\times10^{-7}$ & 27$\pm$2 & T & SGRfig48.png \\
33 & 980712  & 21:50:40.106 & 0.10 & 3.2$\times10^{-5}$ & 8.1$\times10^{-7}$ & 26$\pm$2 & T& 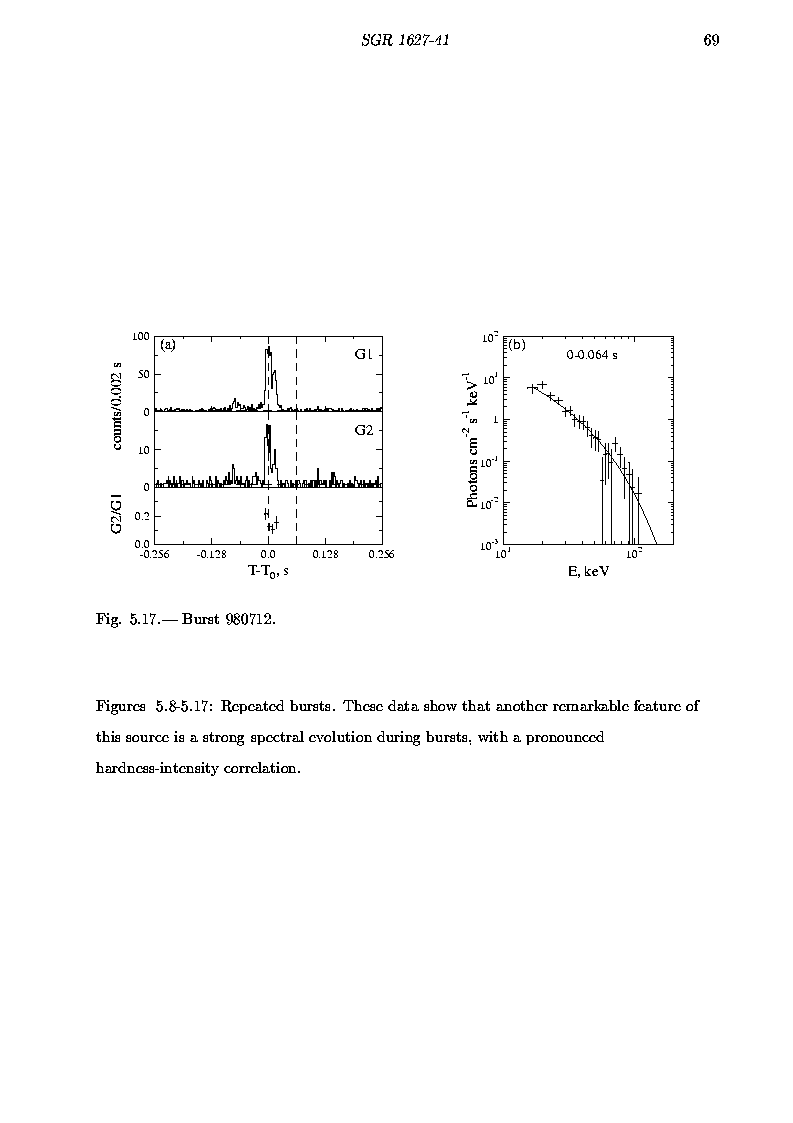 \\
\enddata
\tablenotetext{a}{It is clear that the appearance rate of
soft bursts from this source was very high. 
About 30 events occured during three days in June 1998. 
This explains why about 25 per cent of events were detected in housekeeping data.
A huge outburst was observed on 1998 June 18 (980618d).
At least one more such outburst was detected in housekeeping records,
permitting only a lower limit on the fluence to be determined.}
\end{deluxetable}

\setcounter{subsection}{6}
\setcounter{table}{0}
\begin{deluxetable}{lllccccccc}
\tabletypesize{\scriptsize}
\tablecaption{SGR 1801-23}
\tablehead{
\colhead{N} & \colhead{Burst}& \colhead{T$_0$}& \colhead{$\Delta$T} & \colhead{P$_{max}$} &%
\colhead{S} & \colhead{kT} & \colhead{Comments} & \colhead{Figures} \\
& \colhead{name} &\colhead{h:m:s UT}& \colhead{(s)}&\colhead{(erg cm$^{-2}$ s$^{-1}$)} &%
\colhead{(erg cm$^{-2}$)} & \colhead{(keV)} & & \colhead{PNG file}
}
\startdata
1 & 970629a  & 04:00:24     & $\le1.5$  & \nodata & 1$\times10^{-6}$ & 18$\pm$3 & B& \nodata \\
2 & 970629b  & 06:31:33.221 & 0.25 & 5.4$\times10^{-5}$ & 9.6$\times10^{-6}$ & 16$\pm$1 & T& 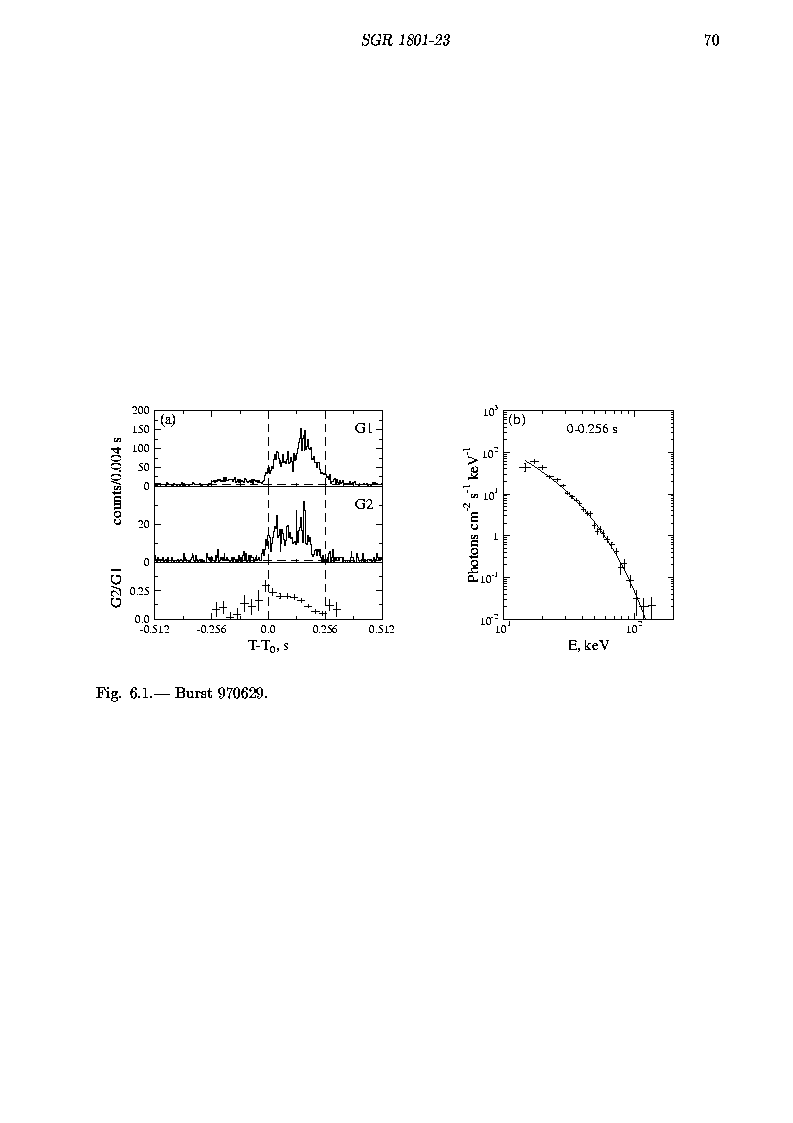 \\
\enddata
\end{deluxetable}


\begin{thebibliography}{}
\bibitem{Aptekar95}Aptekar, R. L., et al. 1995, Space Science Rev., 71, 265
\bibitem{Atteia87}Atteia, J.-L., et al. 1987, \apj, 320, L105
\bibitem{Case98}Case, G. L., \& Bhattacharya, D. 1998, \apj, 504, 761
\bibitem{Cline82}Cline, T. L., et al. 1982, \apj, 255, L45
\bibitem{Cline98}Cline, T. L., Mazets, E. P., \& Golenetskii S. V. 1998,
\iaucirc, 7002
\bibitem{Cline2000}Cline, T. L., et al. 2000, \apj, 531, 407
\bibitem{Dieters98}Dieters, S., et al. 1998, \iaucirc, 6962
\bibitem{Duncan92}Duncan, R. C., \& Thompson, C. 1992, \apj, 392, L9
\bibitem{Ferocy98}Ferocy, M., et al. 1999, \apj, 515, L9
\bibitem{Frederiks97}Frederiks, D. D., et al. 1997, in AIP Conf. Proc. 428,
4th Huntsville Sumposium, Gamma-Ray Bursts, ed. Ch Meegan, R.
Preece, \& T. Koshut (New York: AIP), 921
\bibitem{Gogus99}G\"o\v{g}\"us, E., et al. 1999, \apj, 526, L93
\bibitem{Gogus00}G\"o\v{g}\"us, E., et al. 2000, \apj, 532, L124
\bibitem{Golenetskii74}Golenetskii, S. V., Il'inskii, V. N., \& Mazets, E.
P. 1974, Kosmicheskie Issledovania (Space Research, USSR), 12, 779
\bibitem{Golenetskii84}Golenetskii, S. V., et al. 1984, \nat, 307, 41
\bibitem{Harding99}Harding, A. K., Contopoulos, I., \& Kazans, D. 1999, \apj, 525, L125
\bibitem{Hurley94}Hurley, K., et al. 1994, \apj, 423, 709
\bibitem{Hurley96}Hurley, K., et al. 1996, \apj, 463, L13
\bibitem{Hurley97}Hurley, K., et al. 1997, \iaucirc, 6743
\bibitem{Hurley98}Hurley, K., et al. 1998, \iaucirc, 6929
\bibitem{Hurley99a}Hurley, K., et al. 1999a, \apj, 515, L143
\bibitem{Hurley99b}Hurley, K., et al. 1999b, \apj, 510, L107
\bibitem{Hurley99c}Hurley, K., et al. 1999c, \nat, 397, 41
\bibitem{Hurley99d}Hurley, K., et al. 1999d, \apj, 510, L111
\bibitem{Hurley00}Hurley, K., et al. 2000, \apj, 528, L21
\bibitem{Kouveliotou93}Kouveliotou, C., et al. 1993, \nat, 362, 728
\bibitem{Kouveliotou94}Kouveliotou, C., et al. 1994, \nat, 368, 125
\bibitem{Kouveliotou98}Kouveliotou, C., et al. 1998, \nat, 393, 235
\bibitem{Kouveliotou99}Kouveliotou, C., et al. 1999, \apj, 510, L115
\bibitem{Kulkarni93}Kulkarni, S., \& Frail, D. A. 1993, \nat, 365, 33
\bibitem{Laros86}Laros, J. G., et al. 1986, \nat, 322, 152
\bibitem{Laros87}Laros, J. G., et al. 1987, \apj, 320, L111
\bibitem{Marsden97}Marsden, D., et al. 1997, in AIP Conf. Proc. 428, 4th
Huntsville Symposium, Gamma-Ray Bursts, ed C. Meegan, R.
Preece, \& T. Koshut (New York: AIP), 926
\bibitem{Marsden99}Marsden, D., Rothschild, R. E., \& Lingenfelter, R. E. 1999, \apj, 520, L107
\bibitem{Mazets79b}Mazets, E. P., Golenetskii, S. V., \& Guryan, Yu. A.
1979, Soviet. Astron. Lett., 5, 343
\bibitem{MazetsGol81}Mazets, E. P., \& Golenetskii S. V. 1981, \apss, 75, 47
\bibitem{Mazets79a}Mazets, E. P., et al. 1979a, \nat, 282, 587
\bibitem{Mazets79€}Mazets, E. P., et al. 1979b, Soviet. Astron. Lett., 5, 163
\bibitem{Mazets81}Mazets, E. P., et al. 1981, \apss, 80, 3
\bibitem{Mazets82}Mazets, E. P., et al. 1982, \apss, 84, 173
\bibitem{Mazets83}Mazets, E. P., et al. 1983, in AIP Conf. Proc. 101,
Positron-Electron Pairs in Astrophysics, ed. M. L. Burns, A. K. Harding, \&
R. Ramaty, (New York: AIP), 36
\bibitem{Mazets99a}Mazets, E. P., et al. 1999a, \apj, 519, L151
\bibitem{Mazets99b}Mazets, E. P., et al. 1999b, Astronomy Letters, 10, 727
\bibitem{Mazets99€}Mazets, E. P., et al. 1999c, Astronomy Letters, 10, 735
\bibitem{Murakami84}Murakami, T., et al. 1994, \nat, 368, 127
\bibitem{Norris91}Norris, J. P., et al. 1991, \apj, 366, 240
\bibitem{Rothschild94}Rothschild, R. E., Kulkarni, S. R., \& Lingenfelter, R. E.
1994, \nat, 368, 432
\bibitem{Smith99}Smith, D., et al. 1999, \apj, 519, L147
\bibitem{Thompson95}Thompson, C., \& Duncan, R.C. 1995, \mnras, 275, 255
\bibitem{Vrba00}Vrba, F. J., et al. 2000, \apj, 533, L17
\bibitem{Woods99a}Woods, P. M., et al. 1999a \apj, 519, L139
\bibitem{Woods99b}Woods, P. M., et al. 1999b \apj, 527, L47
\end{thebibliography}
\end{document}